\documentclass[aps,english,showpacs,10pt]{revtex4}

\usepackage[latin1]{inputenc}
\usepackage{float}
\usepackage{latexsym}
\usepackage{graphicx}
\usepackage{amssymb}
\usepackage{amsmath}
\usepackage{amsfonts}
\usepackage{hyperref}
\usepackage{color}
\usepackage{cool}

\begin{document}

\title{The Trans-Planckian Problem in the Healthy Extension of Horava-Lifshitz Gravity}
\author{Elisa G. M. Ferreira}
\email{elisa@fma.if.usp.br/elisafenu@hep.physics.mcgill.ca}
\affiliation{Instituto de Instituto de F\'isica, Universidade de S\~ao Paulo, C.P. 66318,
05315-970, S\~ao Paulo, SP, Brazil and Department of Physics, McGill University, Montr\'eal, QC, H3A 2T8, Canada}
\author{Robert Brandenberger}
\email{rhb@physics.mcgill.ca}
\affiliation{Department of Physics, McGill University, Montr\'eal, QC, H3A 2T8, Canada}
\date{\today}

\begin{abstract}

Planck scale physics may influence the evolution of cosmological fluctuations 
in the early stages of cosmological evolution. Because of the quasi-exponential 
redshifting, which occurs during an inflationary period, the physical wavelengths 
of comoving scales that correspond to the present large-scale structure of the 
Universe were smaller than the Planck length in the early stages of the
inflationary period. This trans-Planckian effect was studied before using
toy models. The Horava-Lifshitz (HL) theory offers the chance to study this problem
in a candidate UV complete theory of gravity. In this paper we study the
evolution of cosmological perturbations according to HL gravity assuming
that matter gives rise to an inflationary background. As is usually done
in inflationary cosmology, we assume that the fluctuations originate in
their minimum energy state. In the trans-Planckian region the fluctuations
obey a non-linear dispersion relation of Corley-Jacobson type. In the 
"healthy extension" of HL gravity there is an extra degree of freedom which
plays an important role in the UV region but decouples in the IR, and which 
influences the cosmological perturbations. We find that in spite of these
important changes compared to the usual description, the overall scale-invariance
of the power spectrum of cosmological perturbations is recovered. However, we obtain
oscillations in the spectrum as a function of wavenumber with a relative
amplitude of order unity and with an effective frequency which scales nonlinearly
with wavenumber. Taking the usual inflationary parameters we find
that the frequency of the oscillations is so large as to render the effect
difficult to observe.

\end{abstract}

\maketitle

\section{Introduction}

The inflationary scenario \cite{Guth} has become the current paradigm to describe 
the early evolution of the universe. The success of this theory comes from a 
number of predictions which have been confirmed by the observations, the main one 
being the causal mechanism for the generation of the primordial cosmological 
perturbation which inflation provides \cite{Mukh}. Inflation is
generally modelled by General Relativity (GR) coupled to scalar field matter. 
The scalar field (called \textit{inflaton}) is responsible for the accelerated 
expansion of space. The success of inflation as a theory for the origin of
structure comes from the fact that the wavelengths of fixed comoving modes which
are of cosmological interest today, namely wavelengths corresponding to the 
present large-scale structure and to the cosmic microwave background anisotropies, 
were sub-Hubble at the beginning of the period of inflation. They cross the 
Hubble horizon about 50 - 60 Hubble times, or \textit{e-folds}, before the end of 
inflation. Thus, inflation must last at least that long to solve the 
problems of the standard cosmological model (SCM) \cite{review_pert,lectures_pert}. 

Most inflationary models have a period of accelerated expansion which lasts 
much more than the 50 - 60 \textit{e-folds} required solve the SCM problems. 
If the period of inflation last for more than about
70 \textit{e-folds} (this number depends very slightly on the energy
scale at which inflation takes place - to get the above number we
are assuming that it is the Grand Unification scale), then all of
the wavelengths important today were smaller that the Planck length at the 
beginning of inflation. In this regime we cannot trust the calculations
since General Relativity will no longer yield a good description of the
physics. This is the known trans-Planckian problem for cosmological fluctuations
\cite{RHBrev0}. A correct computation of both the generation and evolution
of fluctuations in the trans-Planckian regime must be done in the
correct UV completion of the theory.

This problem is named after the analogous trans-Planckian problem for Hawking
radiation in black-hole physics \cite{BHTP}. In the cosmological context this 
problem was discussed in the literature in many different toy model modifications
of General Relativity coupled to a canonical scalar field. These models describe 
the evolution on sub-Planckian wavelengths by imposing \textit{ad-hoc} modifications  
of the usual physics, \textit{e.g.} modified dispersion relations 
\cite{trans_planck,trans_planck_2} or non-commutativity 
\cite{trans_Planck_noncomut} (see also \cite{Amjad}). It is possible to construct toy models in which 
the standard predictions of inflation are changed, in others they are maintained 
\cite{trans_planck} either fully or up to small corrections. For example,
imposing the usual initial conditions on a time-like ``new physics hypersurface'' 
\cite{hypersurface} given by the physical wavelength being some fixed UV cutoff scale
smaller or equal to the Planck scale yields a scale-invariant power
spectrum of curvature fluctuations with small amplitude superimposed
oscillations \footnote{There are also analyses using effective field
theory which discuss limits on the possible magnitude of trans-Planckian
effects \cite{EFT}. Trans-Planckian effects, however, almost by definition go
beyond what can be discussed in an effective field theory approach based
on General Relativity. Bounds on the magnitude of trans-Planckian coming
from demanding that the enhanced fluctuations in the UV do not
destroy the inflationary background are discussed in \cite{BR}.}
If, on the other hand, we assume that a short period of inflation
is preceded by a short non-singular bounce and (before that) the time
reverse of the background evolution, then a deep red spectrum with index 
$n_s = -3$ results \cite{Zhang}. However, to go beyond toy model
studies, we must be able to perform the analysis of cosmological fluctuations
in a full quantum gravity theory. 

Ho\v{r}ava-Lifshitz (HL) gravity \cite{horava} is a candidate theory of
quantum gravity which provides a well motivated framework for describing 
the evolution of the universe on trans-Planckian scales. 
It is an theory of gravity in  $3+1$ dimensions which uses the same
metric degrees of freedom as General Relativity but which is power-counting 
renormalizable (with respect to the scaling symmetry to be introduced below). 
This is achieved by picking a preferred time direction,
thus abandoning the full space-time diffeomorphism invariance (reducing
the symmetry to simply spatial diffeomorphisms), and by introducing an 
anisotropic scaling of Lifshitz type. The loss of space-time isotropy has as 
consequence the absence of Lorentz symmetry, a symmetry which appears as an 
emergent symmetry in the infrared (IR), where GR is recovered.

As a consequence of the loss of the space-dependent time reparametrization
symmetry of General relativity, one of the four gauge symmetries for
cosmological perturbations (namely the space dependent time rescaling
symmetry) is lost, and since the theory presents the same number of basic
degrees of freedom as General Relativity, there is an extra scalar
perturbation mode (scalar in terms of its transformation under spatial
rotations) \cite{extra}.

Since it was proposed by Horava in its most simple version \cite{horava}, 
many other versions of the theory were proposed in the literature
\footnote{For reviews, see \cite{review_HL}}. We have mainly two versions: 
the "projectable", where the lapse function depends only on time to mimic the 
reduced symmetry of the problem; or the "non-projectable" version, where the 
lapse function depends on space and time as in GR. In the "projectable" version 
the extra degree of freedom of the theory is dynamical \cite{RB_ceroni_proj} 
and \textit{ ghost-like} or tachyonic. This version is also plagued by 
"strong coupling" problems \cite{strong_coupling_1,strong_coupling_2}, which
may spoil the validity of this setup. 

In the original form of the "non-projectable" version, the extra degree of 
freedom is non-dynamical \cite{RB_horava} at linear order about a
spatially homogeneous and isotropic background. "Strong coupling" problems 
may arise \cite{strong_coupling_2,strong_coupling_3}. The most studied type 
of a "non-projectable" theory (one which contains all of the terms which
are allowed by power-counting renormalizability and symmetry) is the 
"healthy extension" \cite{healthy} that appears to fix these "strong couplings" 
problems and pathologies associated with the extra degree of freedom that 
appear in the other versions of the theory \footnote{There is still a debate 
in the literature if the  "healthy extension" is indeed free of "strong coupling" 
problems. More on this matter can be found in \cite{strong_coupling_healthy}.}.  
Fluctuations in the healthy extension of HL gravity were first
considered in \cite{Kobayashi}.
As it was shown in \cite{RB_Ceroni_HL}, the extra degree of freedom now
becomes dynamical. With appropriate choices of the parameters in the HL
Lagrangian, the extra degree of freedom is neither ghost-like nor tachyonic. It
is important only in the UV and decouples in the IR. This is interesting 
from the point of view of cosmology since we know, from observations, that 
we can only have one propagating scalar metric degree of freedom in the IR. 
On the other hand, the fact that in the UV there are two propagating degrees of 
freedom makes the theory a promising one to study in the context of the
trans-Planckian problem for fluctuations \footnote{Note that the extra
fluctuating degree of freedom can be eliminated by introducing a local $U(1)$
gauge symmetry \cite{Melby}. Fluctuations in this context have been
analyzed in \cite{Wang}.}.

In the presence of spatial curvature, HL gravity can lead to bouncing
cosmological solutions \cite{RHBbounce}. Here, however, we will assume
the existence of scalar field matter which leads to an inflationary
phase during the evolution of the early universe.
We will use the "healthy extension" of the HL gravity to study the 
trans-Planckian problem for fluctuations. We are interested in checking, 
based on a well motivated description of the UV and IR theory, if the 
predictions of inflation for cosmological perturbations are altered. 
We will have to take into account two effects: the change in the physics 
in the UV region that will be manifest in a modified dispersion relation 
of Corley-Jacobson type \cite{CJ} (as the ones which were studied in 
\cite{trans_planck}); and the presence of an extra degree of freedom that 
will be dynamical in the UV (for wavelengths smaller that Hubble radius in 
our case). This extra degree of freedom can be interpreted as an "entropy" 
perturbation, in analogy with what occurs in multi-field inflation 
models\cite{multifield}, but in this case the extra field is a scalar 
perturbation of the metric. The entropy mode is coupled to the 
"adiabatic perturbation" and will influence its dynamics while the
wavelength is sub-Hubble. Thus, we need to analyse the coupled system of 
differential equations for both scalar perturbations modes in the different 
regimes of the theory in order to evaluate the final power spectrum of the 
"adiabatic" mode.

The article is organized as follows. In Section \ref{sec:setup}, we give a 
brief description of the  "healthy extension" of HL, setting up the theory 
we are using and calculating linear perturbations about a homogeneous and 
isotropic background. We show how the reduced symmetry group gauges only one 
of the degrees of freedom, leaving one extra degree of freedom, by calculating 
the second order action. In Section \ref{sec:cosm_pert} we study the 
cosmological perturbations. First we find the canonical variables of the 
problem and show that one of these corresponds to the usual Mukhanov-Sasaki 
variable and the other decouples in the IR. We take the UV limit of the theory 
to find the contributions of the trans-Planckian physics. The initial state 
is calculated by minimizing the vacuum energy and the solutions to the 
inhomogeneous equations of motion are found in each region of the problem. 
Finally, the power spectrum is calculated. We find that the overall
scale-invariance of the spectrum is maintained, but that there are
superimposed oscillations with an effective frequency depending on the
wavenumber (very different from the oscillations which are seen in
the approach of \cite{hypersurface} to the trans-Planckian problem). The amplitude
of these oscillations is not suppressed (again in contrast to what is
seen in \cite{hypersurface}). However, the frequency of oscillation is very large
making the effect extremely difficult to measure.

\section{Health Extension of Horava-Lifshitz}
\label{sec:setup}

\subsection{Background}

In HL theory it is natural to use the ADM (Arnowitt, Desner and Misner) formalism 
that separates space-time into time and spatial foliations:
\begin{equation}
ds^{2}=-N^{2}dt^{2}+g_{ij}\left(dx^{i}-N^{i}dt\right)\left(dx^{j}-N^{j}dt\right)\,,
\end{equation}
where $g_{ij}\left(t,\, x\right)$ is the metric of the spatial section, 
\textit{N} is the lapse function and  $N^{i}\left(t,\, x\right)$ is the shift 
vector. As mentioned, the lapse function can be restricted to depend only on 
time, yielding the "projectable" theory, or else it is taken to depend on both
space and time, yielding the "non-projectable" version. The "healthy extension" 
is a non-projectable version of the theory with all of the terms allowed
by the residual symmetries and by power-counting renormalizability included.
Thus $N(t,x)$ and we describe the same lapse function as in GR. In this
case, we must include in the action terms proportional to the quantity:
\begin{equation}
a_{i}=\frac{\partial _{i}N}{N}\,.
\end{equation}
Because of the anisotropic scaling, the classical scaling dimensions of the 
coordinates of the theory (when the scaling dimension is $z=3$ as it is
in four space-time-dimensional HL gravity) are, in units of spatial momenta: 
$[t]_{s} = -3$, $[x_{i}]_{s} = -1$. With this we have that 
$[\nabla \equiv \partial _{i} \partial ^{i}] = 2$. The ADM fields have scaling 
dimension: $[g_{ij}]_{s}=[N]_{s}=0$ and $[N_{i}]$=2.

The action of HL gravity is constructed by including all terms with scaling 
dimension smaller or equal to six, in order to obtain a theory which is
power-counting renormalizable, and which are invariant under the 
spatial diffeomorphisms, i.e. under elements of the symmetry
group $Diff_{\mathcal{F}}$. The "healthy extension" action can be written as, 
following  \cite{RB_Ceroni_HL} (see also \cite{wang_maartens,wang_wands_maartens}):
\begin{equation}
S=\chi ^{2} \int dt d^{3} x \sqrt{g} N \left( \mathcal{L}_{kin}-\mathcal{L}_{V}-\mathcal{L}_{E}+\chi ^{-2} \mathcal{L}_{M} \right)\,,
\end{equation}
where $\chi ^{2}=1/16\pi G$. The kinetic term is the one that contains time 
derivatives of the metric in a covariant way with respect to the new 
foliating preserving symmetry, $Diff_{\mathcal{F}}$:
\begin{equation}
\mathcal{L}_{kin}=K_{ij}K^{ij}-\lambda K^{2}\,,
\end{equation}
where $K_{ij} =(1/2N)\left(\dot{g}_{ij}-\nabla _{i}N_{j}-\nabla _{j} N_{i}\right)$ 
is the extrinsic curvature of the constant time hypersurfaces. The potential term 
is constructed by including all terms that preserve $Diff_{\mathcal{F}}$ until the 
sixth spatial derivative:
 \begin{alignat}{1}
\mathcal{L}_{V}&=2\Lambda -R+\frac{1}{\chi ^{2}} \left( g_{2}R^{2}+g^{3}R_{ij}R^{ij} \right)+\frac{1}{\chi ^{4}} \left[g_{4} R^{3}+g_{5}R R_{ij}R^{ij}+g_{6}R^{i}_{j}R^{j}_{k}R^{k}_{i} \right. \nonumber \\
&+\left. g_{7} R \nabla ^{2} R+g_{8} \left( \nabla _{i} R_{jk} \right) \left( \nabla ^{i} R^{jk} \right) \right]\,,
\end{alignat}
where $\Lambda$ is the cosmological constant and has dimension of momentum 
squared and the coefficients are dimensionless. The new components proportional 
to $a_{i}$ that appear in the "healthy extension" are:
\begin{alignat}{1}
\mathcal{L}_{E}=-\eta a_{i}a^{i} +\frac{1}{\chi ^{2}} \left( \eta_{2}a_{i} \triangle a^{i}+\eta_{3}R\nabla_{i}a^{i}+\ldots\right)
&+\frac{1}{\chi ^{4}}\left[\eta_{4}a_{i}\triangle ^{2}a^{i}+\eta _{5}\triangle R \nabla_{i}a^{i}+
\eta_ {6}R^{2} \nabla_{i}a^{i}+\ldots \right]\,,\label{he}
\end{alignat}
where the coefficients are dimensionless. As we are interested in obtaining  
an inflationary background cosmology, we include matter in the form of a scalar field, 
with matter Lagrangian \cite{lifshitz}
\begin{equation}
\mathcal{L}_{M}=\frac{1}{2N^{2}}\left(\dot{\phi}-N^{i}\partial_{i}\phi\right)^{2}+V\left(\phi,\,\partial_{i}\phi,\, g_{ij}\right)\,,\label{eq:escala_HL_1}
\end{equation}
where the potential terms are given by:
\begin{equation}
V\left(\phi,\,\partial_{i}\phi,\, g_{ij}\right)=V_{0}+V_{1}P_{0}+V_{2}P_{1}^{2}+V_{3}P_{1}^{3}+V_{4}\phi P_{2}+V_{5}P_{0}P{2}+V_{6}P_{1}P_{2}\,,
\end{equation}
with $P_{0}\equiv \left(\nabla \phi \right) ^{2}$, 
$P_{i}\equiv \triangle ^{i} \phi$ and 
$\triangle \equiv g^{ij}\nabla _{i} \nabla _{j}$. 
The scalar field has scaling dimension $[\phi ]_{s}=(d-z)/2=0$. In the way it is 
written above, the coefficients are not dimensionless as the previous coefficients
are. If we want to write them as dimensionless coefficients we have to make the 
following substitution: the order 2 coefficient $V_{1}$  is already dimensionless; 
the order four coefficients are changed to 
$V_{2} \rightarrow \tilde{V}_{2} /\chi ^{2}$ and 
$V_{4} \rightarrow \tilde{V}_{4} /\chi ^{2}$, 
with $\tilde{V}_{2}$ and $\tilde{V}_{4}$ dimensionless; and the 
dimension six coefficients are changed to 
$V_{3} \rightarrow \tilde{V}_{3} /\chi ^{4}$, 
$V_{5} \rightarrow \tilde{V}_{5} /\chi ^{4}$ and 
$V_{6} \rightarrow \tilde{V}_{6} /\chi ^{4}$, 
with $\tilde{V}_{3}$, $\tilde{V}_{5}$ and $\tilde{V}_{6}$ dimensionless.

The non-projectable version of HL can contain more than 60 terms. Here we only 
considered the ones that are important for the scalar part of the perturbations, 
the perturbations we interested in. In the IR the GR action is recovered when 
$\lambda =1$.

We can now obtain the Hamiltonian and momentum constraints. Varying the 
action with respect to  $N^{i}\left(t,\, x^{j}\right)$, we have the momentum 
constraint:
\begin{equation} 
\nabla_{i}\pi^{ij}= \frac{\kappa^{2}}{4}J^{j}\,, 
\label{eq:vinc_momentum}
\end{equation}
where 
$\pi ^{ij} \equiv\frac{\delta\mathcal{L}_{kin}}{\delta\dot{g}_{ij}}=-K^{ij}+\lambda K g^{ij}$ 
and 
$J_{i}  \equiv -N\frac{\delta\mathcal{L}_{\phi}}{\delta N^{i}}=\frac{1}{N}\left(\dot{\phi}-N^{k}\nabla_{k}\phi\right)\nabla_{i}\phi$. 
Note that (\ref{eq:vinc_momentum}) is a local constraint.

The Hamiltonian constraint can be obtained by varying the action with respect 
to the lapse function, $N(t,x^{i})$: 
\begin{equation}
\mathcal{L}_{cin}+\mathcal{L}_{pot}+\mathcal{L}_{E}+N\frac{\delta\mathcal{L}_{E}}{\delta N}=\frac{1}{4\chi ^{2}}J^{t}\,, \label{hamiltonian_const}
\end{equation}
with 
$J^{t} \equiv 2\left(N\frac{\delta\mathcal{L}_{\phi}}{\delta N}+\mathcal{L}_{\phi}\right)$
and the new terms from "healthy extension":
\begin{equation}
 N\frac{\delta\mathcal{L}_{E}}{\delta N}=2\eta \nabla _{i}a^{i}-\frac{2\eta _{2}}{\chi ^{2}} \triangle \nabla _{i}a^{i}+\frac{\eta _{3}}{\chi ^{2}} \triangle R-\frac{2\eta _{4}}{\chi ^{4}} \triangle ^{2} \nabla _{i}a^{i}+\frac{\eta _{5}}{\chi ^{4}}\triangle ^{2} R+\frac{\eta _{6}}{\chi ^{6}} \triangle R^{2}+...\,. \label{hamiltonian_const_2}
\end{equation}
This constraint is a local constraint, unlike what happens in the "projectable" 
version of the theory. 

One might think that the presence of two local constraints should lead to a 
reduction of the number of degrees of freedom of the theory to have the same 
number as in General Relativity. However, this is not the case. The local 
Hamiltonian constraint is not able to decouple the extra degree of freedom in 
all the regimes of the theory, only in the IR. We can see this by studying 
cosmological perturbation theory.

\subsection{Perturbations}

Perturbing about an isotropic and homogeneous universe described by the 
spatially flat FRW metric, 
\begin{equation}
ds^{2} = -dt^{2}+a^{2}\delta _{ij}dx^{i}dx^{j} \, ,
\end{equation}
the background values and scalar perturbation of the ADM variables are given 
by \cite{review_pert}:
\begin{alignat}{1}
N+\delta N\left(t,\, x^{k}\right) & =1+\nu\left(t,\, x^{k}\right)\,,\\
N_{i}+\delta N_{i}\left(t,\, x^{k}\right) & =\partial_{i}B\left(t,\, x^{k}\right)\,,\\
g_{ij}+\delta g_{ij}\left(t,\, x^{k}\right) & =a^{2}\delta _{ij}+a^{2}\left(t\right)\left[-2\psi\left(t,\, x^{k}\right)\delta_{ij}+2E\left(t,\, x^{k}\right)_{;ij}\right]\,.
\end{alignat}
We must also consider the perturbation in the scalar field, given by: 
\begin{equation}
\phi \left(t,\, x^{k}\right)=\phi_{0}\left(t\right)+\delta\phi\left(t,\, x^{k}\right)\,.
\end{equation} 

The invariance under $Diff_{\mathcal{F}}$, allow us to set $E=0$.  Because of the 
reduced symmetry we can no longer constrain another degree of freedom of the 
perturbations and use the standard gauges used in GR \cite{review_pert}. Thus 
we are left with more degrees of freedom than we would have in GR:  
$\nu$, $B$, $\psi$ and $\delta\phi$. We can reduce the number of them by 
solving the constraints of the theory. 

Expanding the momentum and Hamiltonian constraints to first order, we can use 
them to solve only for two degrees of freedom (in a spatially flat background), 
namely $\nu$ and $B$. In the momentum space, we can rewrite the equation 
using the physical momentum $\bar{k}\equiv k/a$ (as done in \cite{RB_Ceroni_HL}): 
\begin{alignat}{1}
d\left(k\right)B_{k}\left(t\right) & =-\left(3\lambda-1\right)\left[\frac{\dot{\phi}_{0}^{2}}{\chi ^{2}\bar{k}^{2}}+2f_{1}\left(\bar{k}\right)\right]\dot{\psi}_{k}\left(t\right)-\left(3\lambda-1\right)\frac{H\dot{\phi}_{0}^{2}}{\chi ^{2}\bar{k}^{2}}\dot{\delta\phi_{k}}\left(t\right)-4\left(3\lambda-1\right)f_{2}\left(\bar{k}\right)H\psi_{k}\left(t\right)-\nonumber \\
 & -\left\{ \left(3\lambda-1\right)\left[V_{o,\phi}\left(\phi_{0}\right)+V_{4}\left(\phi_{0}\right)\bar{k}^{4}H+3\left(3\lambda-1\right)\dot{\phi}_{0}H^{2}-\frac{\dot{\phi}_{0}^{3}}{2\chi ^{2}}-\dot{\phi}_{0}f_{1}\left(\bar{k}\right)\bar{k}^{2}\right]\right\} \frac{\delta\phi_{k}\left(t\right)}{\chi ^{2}\bar{k}^{2}}\,, \\ 
 d\left(\bar{k}\right)\nu_{k}\left(t\right) & =\left(\lambda-1\right)\frac{\dot{\phi}_{0}}{\chi ^{2}}\dot{\delta\phi_{k}}\left(t\right)-4\left(3\lambda-1\right)H\dot{\psi}_{k}\left(t\right)+4\left(\lambda-1\right)f_{2}\left(\bar{k}\right)\bar{k}^{2}\psi_{k}\left(t\right)+\nonumber \\
 & +\left\{ \left(3\lambda-1\right)\phi_{0}H+\left(\lambda-1\right)\left[V_{0,\phi}\left(\phi_{0}\right)+V_{4}\left(\phi_{0}\right)\bar{k}^{4}\right]\right\} \frac{\delta\phi_{k}\left(t\right)}{\chi ^{2}}\,,\label{eq:const_HE}
\end{alignat}
where we introduced the following functions to simplify these equations:
\begin{align}
f_{1}\left(\bar{k}\right) & \equiv-\eta+\eta_{2}\frac{\bar{k}^{2}}{\chi ^{2}}+\eta_{4}\frac{\bar{k}^{4}}{\chi ^{4}}\,,\qquad
f_{2}\left(\bar{k}\right)\equiv-1+\eta_{3}\frac{\bar{k}^{2}}{\chi ^{2}}+\eta_{5}\frac{\bar{k}^{4}}{\chi ^{4}}\,,\\
d\left(\bar{k}\right) & \equiv 4\left(3\lambda-1\right)H^{2}+\left(\lambda-1\right)\left[\frac{\dot{\phi}_{0}^{2}}{\chi ^{2}}+2f_{1}\left(\bar{k}\right)\bar{k}^{2}\right]\\
 & =4\left(3\lambda-1\right)H^{2}\left[1+\frac{\lambda-1}{2\left(3\lambda-1\right)}\frac{\dot{\phi}_{0}^{2}}{\chi ^{2}}H^{2}+\frac{\lambda-1}{2\left(3\lambda-1\right)}\left(-\eta+\eta_{2}\frac{\bar{k}^{2}}{\chi ^{2}}+\eta_{4}\frac{\bar{k}^{4}}{\chi ^{2}}\right)\frac{\bar{k}^{2}}{H^{2}}\right]\,,\label{eq:d_limite}
\end{align}
Since we have the function $d\left(\bar{k}\right)$ in both solutions, 
$B_{k}\left(t\right)$ and $\nu_{k}\left(t\right)$ are regular in the limit 
$\lambda\rightarrow 1$ as long as $H \neq 0$.
The second expression in (\ref{eq:d_limite}) is valid only when $\lambda\neq1/3$ 
and $H\neq0$. From this, we can see that the value of the physical momentum that 
separates the regions of high and low energy is proportional to $H$, the square
of the corresponding length being
$\frac{\lambda-1}{2\left(3\lambda-1\right)} \frac{1}{H^{2}}$. We are going to see 
later that in the de Sitter case this corresponds exactly to the Hubble horizon, the 
scale where the cosmological perturbations freeze out. Thus, the scale that separates 
the UV and IR limits in the "healthy extension" of HL gravity it is the same as the 
event horizon of de Sitter inflation. Beyond this length scale the evolution is similar 
to GR.

We can write the second order Lagrangian for the perturbations using the constraints 
(\ref{eq:const_HE}) in the following way:
\begin{alignat}{1}
\delta_{2}S_{scalar} & =\chi ^{2}\intop dt\frac{d^{3}k}{\left(2\pi\right)^{3}}a^{3}(c_{\phi}\dot{\delta\phi_{k}^{2}}+c_{\psi}\dot{\psi}_{k}^{2}+c_{\phi\psi}\dot{\delta\phi_{k}^{2}}\dot{\psi}_{k}^{2}+f_{\phi}\delta\phi_{k}\dot{\delta\phi_{k}}+f_{\psi}\psi_{k}\dot{\psi}_{k}\nonumber \\
 & +f_{\phi\psi}\psi_{k}\dot{\delta\phi_{k}}+\bar{f}_{\phi\psi}\dot{\psi}_{k}\delta\phi_{k}-m_{\phi}^{2}\delta\phi_{k}^{2}-m_{\psi}^{2}\psi_{k}^{2}-m_{\phi\psi}^{2}\psi_{k}\delta\phi_{k})\,, \label{S_2ordem}
\end{alignat}
where the coefficients are given in Appendix 1. We can see that this action has 
two dynamical degrees of freedom with mixed kinetic terms. They cannot be combined 
into one degree of freedom, the Mukhanov-Sasaki variable, 
$-\zeta \equiv \psi + \frac{H}{\dot{\phi_{0}}}\delta\phi$, 
as is done in the case of cosmological perturbations in GR 
\cite{review_pert,lectures_pert} or even in 
the "non-projectable" version of HL gravity with detailed balance \cite{RB_horava}. 

We cannot diagonalize the entire action. Similar to what it is done in multi-field 
inflation, we are going to make a rotation in field space to find the true propagating 
and canonical variables of the problem 
\footnote{Since we know from the theory of cosmological perturbations in GR that 
the curvature fluctuations at late times are given by a combination of metric and 
matter perturbations, we expect that also here the canonical variables will be a 
combination of the matter perturbation $\delta \phi $ and the scalar perturbation 
$\psi$ of the metric .}. For this we will
diagonalize the kinetic term of the action. We can write the action in vector and 
matrix form as:
\begin{alignat}{1}
\delta_{2}S & =\chi ^{2}\int dt\frac{d^{3}k}{(2\pi)^{3}}a^{3}\left[\left(\begin{array}{cc}
\dot{\delta\phi_{k}} & \dot{\psi_{k}}\end{array}\right)\left(\begin{array}{cc}
c_{\phi} & \frac{1}{2}c_{\phi \psi}\\
\frac{1}{2}c_{\phi \psi} & c_{\psi}
\end{array}\right)\left(\begin{array}{c}
\dot{\delta\phi_{k}}\\
\dot{\psi_{k}}
\end{array}\right)+\left(\begin{array}{cc}
\dot{\delta\phi_{k}} & \dot{\psi_{k}}\end{array}\right)\left(\begin{array}{cc}
f_{\phi} & f_{\phi\psi}\\
\tilde{f_{\phi\psi}} & f_{\psi}
\end{array}\right)\left(\begin{array}{c}
\delta\phi_{k}\\
\psi_{k}
\end{array}\right)\right.\nonumber \\
 & \left.+\left(\begin{array}{cc}
\delta\phi_{k} & \psi_{k}\end{array}\right)\left(\begin{array}{cc}
-m_{\phi}^{2} & -\frac{1}{2}m_{\phi\psi}^{2}\\
-\frac{1}{2}m_{\phi\psi}^{2} & -m_{\psi}^{2}
\end{array}\right)\left(\begin{array}{c}
\delta\phi_{k}\\
\psi_{k}
\end{array}\right)\right]\,,
\end{alignat}
Diagonalizing the kinetic term which we call $\mathcal{C}$, we find that the 
eigenvalues are given by \footnote{We did not rescale the variable $\phi _{k}$ 
as was done in \cite{RB_Ceroni_HL}, since 
without the rescaling the observables have the right dimension, as will be seen 
in the following section.}:
\begin{alignat}{1}
d\left(\bar{k}\right)\lambda _{1,2} & =\frac{\left(3\lambda -1\right)}{\chi ^{2}} \left\{ \left(H^{2}+\dot{\phi}_{0}^{2} \right) +f_{1}\left(\bar{k}\right)\bar{k}^{2} \left[ \frac{\left( \lambda -1	 \right)}{2 \left( 3\lambda -1 \right) } +2\chi ^{2} \right] \right\} \nonumber \\
 & \pm\sqrt{\frac{\left(3\lambda -1\right)^{2}}{\chi ^{4}} \left\{ \left(H^{2}+\dot{\phi}_{0}^{2} \right)  ^{2}+2\left(H^{2}-\dot{\phi}_{0}^{2} \right)  \left[ \frac{\left( \lambda -1	 \right)}{2 \left( 3\lambda -1 \right) } -2\chi ^{2} \right] +f_{1} ^{2}\left(\bar{k}\right)\bar{k}^{4} \left[ \frac{\left( \lambda -1 \right)}{2 \left( 3\lambda -1 \right) } -2\chi ^{2} \right]^{2} 
 \right\} }\,.\label{autoval}
\end{alignat}
Given these eingenvalues we have to determine the rotation matrix $S$ that 
diagonalizes $\mathcal{C}$ and then apply it to the entire action.  Imposing the 
normalization condition, the eigenvectors of $\mathcal{C}$ can be determined and 
the rotation angle can be inferred from:
\begin{equation}
\tan (2\theta )=\frac{c_{\phi\psi}}{c_{\psi}-c_{\phi}}\,.
\end{equation} 
We need to apply the diagonalization matrix in the entire action to write it in 
terms of the new variables. As a result of this algebra we find that, up to total 
derivative terms, we have that the action in terms
of the new basis with diagonalized kinetic term is given by:
\begin{alignat}{1}
\delta _{2} S=\int dt\frac{d^{3}k}{(2\pi)^{3}}a^{3}\left[c_{1} \dot{\phi}^{2}_{3}+c_{2} \dot{\phi}^{2}_{4}+F_{\phi _{3}}\dot{\phi}_{3} \phi _{3}+F_{\phi _{4}}\dot{\phi}_{4} \phi _{4}+F_{\phi _{3} \phi _{4}} \dot{\phi}_{3} \phi _{4}+\tilde{F}_{\phi _{3} \phi _{4}} \phi _{3} \dot{\phi}_{4}-M^{2}_{\phi _{3}} \phi _{3}^{2}-M^{2}_{\phi _{4}} \phi _{4}^{2}-M^{2}_{\phi _{3}\phi _{4}} \phi _{3} \phi _{4} \right] \,.
\end{alignat}
where the new rotated variables are given by 
\begin{equation}
\left( \begin{array}{c} \phi _{3} \\ \phi _{4} \end{array} \right)=S^{t} c_{2}
\end{equation} 
and the new coefficients are given in Appendix 2.

\section{Cosmological Perturbations and the Power Spectrum}
\label{sec:cosm_pert}

Because of the power-counting renormalizability, HL gravity can contain terms up to 
sixth order in spatial derivatives. The terms of highest order in momentum dominate 
the evolution in the UV. As pointed out in \cite{inv_spectrum_HL_1,inv_spectrum_HL_2} 
this will lead to a scale invariant initial power spectrum of vacuum fluctuations. The 
reason is that the form of the state which minimizes the HL Hamiltonian is different 
from the one which minimizes the usual Hamiltonian of a Lorentz-invariant theory. The 
difference in Hamiltonians also affects the evolution of the state. We have to 
investigate the evolution of the perturbations that are generated inside the trans-Planck 
region until the IR region where we recover the GR evolution and verify how the new 
physics in the UV modifies the power spectrum.

Another effect in the "healthy extension" is the presence of an extra degree of freedom 
that will only be dynamical in the UV region of the theory, inside the Hubble radius. 
The coupling of this extra mode to the adiabatic mode can also be a source of 
modifications of the power spectrum of the adiabatic fluctuation mode. In the following 
we will investigate these effects.

\subsection{Equations of Motion}

It is convenient to perform a change of variables to absorb the multiplicative factors 
in front the kinetic term and absorb the scale factor. The rescaled variables are: 
\begin{eqnarray}
u_{k} = a\sqrt{\lambda _{1}} \phi _{3}\,, \qquad
v_{k} = a\sqrt{\lambda _{2}} \phi _{4} \, .  
\end{eqnarray}
Up to terms which are total derivatives, the action becomes:
\begin{equation}
\delta _{2} S=\chi ^{2} \int d\eta \frac{d^{3}k}{(2\pi )^{3}} \left[u_{k}^{'2}+v_{k}^{'2}+\frac{F_{\phi _{3} \phi _{4}}a}{\sqrt{\lambda _{1} \lambda _{2}}}u_{k}^{'}v_{k}+\frac{\tilde{F}_{\phi _{3} \phi _{4}}a}{\sqrt{\lambda _{1} \lambda _{2}}}u_{k} v_{k}^{'}+\omega ^{2}_{u} u_{k}^{2}+\omega ^{2}_{v} v_{k}^{2}-M^{2}_{uv} u_{k} v_{k} \right]\,, \label{S2_conforme}
\end{equation}
where:
\begin{alignat}{1}
\omega ^{2}_{u}=a^{2} \frac{3HF_{\phi _{3}}+\dot{F}_{\phi _{3}}-M^{2}_{\phi _{3}} }{\lambda _{1}}+\left( \mathcal{H}+\frac{1}{2} \frac{\lambda _{1}^{'}}{\lambda _{1}} \right) ^{'}+\left( \mathcal{H}+\frac{1}{2} \frac{\lambda _{1}^{'}}{\lambda _{1}} \right)^{2} \,, \\
\omega ^{2}_{v}=a^{2} \frac{3HF_{\phi _{4}}+\dot{F}_{\phi _{4}}-M^{2}_{\phi _{4}} }{\lambda _{2}}+\left( \mathcal{H}+\frac{1}{2} \frac{\lambda _{2}^{'}}{\lambda _{2}} \right) ^{'}+\left( \mathcal{H}+\frac{1}{2} \frac{\lambda _{2}^{'}}{\lambda _{2}} \right)^{2} \,, \\
-M^{2}_{uv}=\frac{1}{\sqrt{\lambda _{1} \lambda _{2}}} \left[-F_{\phi _{3} \phi _{4}} \, a \left( \mathcal{H}+\frac{1}{2} \frac{\lambda _{1}^{'}}{\lambda _{1}} \right)-\tilde{F}_{\phi _{3} \phi _{4}} \, a \left( \mathcal{H}+\frac{1}{2} \frac{\lambda _{2}^{'}}{\lambda _{2}} \right)-M^{2}_{\phi _{3} \phi _{4}} a^{2} \right]\,. \label{coef_u_v}
\end{alignat}
With this, we can derive the equations of motion for both dynamical variables of the problem:
\begin{equation}
u_{k}^{''}-\omega ^{2}_{u} u_{k}=-\frac{1}{2} \{ v_{k}^{'} a\frac{\left( F_{\phi _{3} \phi _{4}}-\tilde{F}_{\phi _{3} \phi _{4}} \right)}{\sqrt{\lambda _{1} \lambda _{2}}}+v_{k} \left[M^{2}_{uv}+\left( \frac{F_{\phi _{3} \phi _{4}}a}{\sqrt{\lambda _{1} \lambda _{2}}} \right) ^{'}\right] \}\,, \label{eom_u}
\end{equation}
\begin{equation}
v_{k}^{''}-\omega ^{2}_{v} v_{k}=-\frac{1}{2} \{ u_{k}^{'} a \frac{\left( \tilde{F}_{\phi _{3} \phi _{4}} -F_{\phi _{3} \phi _{4}} \right)}{\sqrt{\lambda _{1} \lambda _{2}}}+u_{k} \left[M^{2}_{uv}+\left( \frac{\tilde{F}_{\phi _{3} \phi _{4}}a}{\sqrt{\lambda _{1} \lambda _{2}}} \right) ^{'}\right] \}\,. \label{eom_v}
\end{equation}
The evolution of these variables is of the form of coupled oscillators with time 
dependent mass, where the right side of the equations represents the interaction 
between the oscillators.

After these calculations, some consistency checks must be done to see if the system 
describes the expected physics:

\begin{itemize}
\item \textit{First consistency check:} We need to check if the variable $\phi _{3}$ is the 
Mukhanov-Sasaki variable, the one that we want to be the right canonical variable described 
by the perturbations in the IR. In order to check this, lets evaluate 
$\phi _{3}$ e $\phi _{4}$:
\begin{eqnarray}
\phi ^{2}_{3}=\frac{1}{2}(1+\cos 2\theta ) \delta \phi _{k}^{2}+\frac{1}{2}(1-\cos 2\theta )  \psi _{k}^{2}+\sin 2\theta \, \delta \phi _{k}\psi _{k} \,,\\ 
\phi _{4}^{2}=\frac{1}{2}(1-\cos 2\theta ) \delta \phi _{k}^{2}+\frac{1}{2}(1+\cos 2\theta )  \psi _{k}^{2}-\sin 2\theta \, \delta \phi _{k} \psi _{k} \,,
\end{eqnarray}
and using (\ref{sin_cos}) we can see that in the IR and when 
$H^{2} \gg \dot{\phi}^{2}_{0}/\chi^{2}$ we have:
\begin{eqnarray}
\phi_{3} \sim \delta \phi _{k}+\frac{H}{\dot{\phi _{0}}}\psi _{k}\,,\,\,\,\,\,\, \phi_ {4} \sim \frac{H}{\dot{\phi _{0}}} \delta \phi _{k}- \psi _{k} \,.
\end{eqnarray}
We can see that the new variable $\phi _{3}$ has the form of the usual Mukhanov-Sasaki 
variable in the IR and it will hence describe the right observable.

\item \textit{Second consistency check:} Now, we need to verify that the extra 
variable $\phi _{4}$ decouples in the IR, which means that its mass diverges in this 
limit, which then leads to the presence of a single propagating degree of freedom. 
Considering  $H^{2} \gg \dot{\phi}^{2}_{0}$, we obtain that in the IR the eigenvalues are:
\begin{eqnarray}
\lambda _{1} \rightarrow \frac{1}{2\chi ^{2}}\,,\,\,\,\,\,\,\, \lambda _{2} \rightarrow -\eta \left(\frac{k}{aH} \right) ^{2}\,, \label{av_IR}
\end{eqnarray}
and in the UV:
\begin{eqnarray}
\lambda _{1} \rightarrow \frac{1}{2\chi ^{2}}\,,\,\,\,\,\,\,\, \lambda _{2} \rightarrow 2\frac{3\lambda -1}{\lambda -1} \, . \label{av_UV}
\end{eqnarray}

From (\ref{eom_u}), (\ref{eom_v}) and (\ref{coef_u_v}) we can see that the mass terms 
of the equations are divided by the respective eigenvalue. So, from (\ref{av_IR}), we 
infer that  $\lambda _{2}$ goes to zero in the IR when $k \rightarrow 0$  and its mass 
goes to infinity. The extra degree of freedom  $\phi _{4}$ decouples in this limit, 
as required for the theory to have an IR behaviour compatible with observations.
\end{itemize}

\vspace{0.5cm}

Having verified the consistency relations we need to evaluate the coefficients appearing 
in the equations of motion. We want to study the dynamics in the UV limit, as we are 
interested in investigating the trans-Planckian problem and the power spectrum at Hubble 
horizon exit. In this way, we will determine the coefficients in this regime, using the 
coefficients of Appendix 2 and (\ref{av_IR},\ref{av_UV}). 

\subsection{UV Limit}

In the UV, the most important contributions come from the highest orders in $k$. 
Therefore we will only consider the highest order contributions in each coefficient:
\begin{eqnarray}
F_{\phi _{3}} &\propto & \mathcal{O}(1/k^{2})\,; \hspace{1cm} F_{\phi _{3} \phi _{4}}\propto \mathcal{O}(const.)\,;\hspace{1cm} M^{2}_{\phi _{3}} \propto \mathcal{O}(k^{6})\,;\hspace{1cm} M^{2}_{\phi _{3} \phi _{4}}\propto \mathcal{O}(k^{4})\,;\\
F_{\phi _{4}} &\propto & \mathcal{O}(1/k^{2})\,; \hspace{1cm}\tilde{F}_{\phi _{3} \phi _{4}} \propto \mathcal{O}(const.)\,;\hspace{1cm}M^{2}_{\phi _{4}} \propto \mathcal{O}(k^{6})\,.
\end{eqnarray}
The contributions of $F_{\phi _{3}}$ and $F_{\phi _{4}}$ are negligible in 
comparison with $M^{2}_{\phi _{3}}$ and $M^{2}_{\phi _{4}}$, which are also present 
in $\omega ^{2}_{u}$ and in  $\omega ^{2}_{v}$, respectively. Hence we can simplify
the equations of motion in the UV by considering 
$\omega ^{2}_{u} \sim -2M^{2}_{\phi _{3}}+\mathcal{H}^{'}+\mathcal{H}^{2}$ and 
$\omega ^{2}_{v} \sim -2M^{2}_{\phi _{4}}+\mathcal{H}^{'}+\mathcal{H}^{2}$:
\begin{eqnarray}
u_{k}^{''}+\left( \frac{\alpha ^{2}k^{6}}{a^{4}}+\frac{\beta ^{2}k^{4}}{a^{2}}+ c_{s}^{2}k^{2}-\mathcal{H}^{'}-\mathcal{H}^{2} \right) u_{k}=-v_{k}^{'} a \Xi _{2}+v_{k} \left( \frac{\gamma ^{2}k^{4}}{a^{2}}+\Xi _{3} a^{2} \right) \,, \\
v_{k}^{''}+\left( \frac{\delta ^{2}k^{6}}{a^{4}}+\frac{\epsilon ^{2}k^{4}}{a^{2}}+ \tilde{c}_{s}^{2}k^{2}-\mathcal{H}^{'}-\mathcal{H}^{2} \right) v_{k}=-u_{k}^{'} a \Omega _{2}+u_{k} \left( \frac{\zeta ^{2}k^{4}}{a^{2}}+\Omega _{3} a^{2} \right) \,. \label{eom_uv}
\end{eqnarray}
where in the interaction terms we also considered the constant terms (which are 
the leading ones) of $F_{\phi _{3} \phi _{4}}$ and $\tilde{F}_{\phi _{3} \phi _{4}}$ since 
they are accompanied by a time derivative term. Without a careful analysis of the 
influence of these terms we cannot neglect them. The remaining coefficients are given 
in Appendix 3 in terms of the dimensionless coefficients. We can see that the terms 
have the expected scaling dimension since we know that $[\omega ^{2}]_{s}=2$.  

The coefficient functions of the $u_k$ and $v_k$ terms on the left hand side of
the above equations determine the dispersion relations of the mode functions
in the UV. The dominant term is the $k^6$ term. These dispersion relations encode 
the change in the physics in the UV due to the presence of the higher space derivative
terms in the HL action.  These dispersion relations are of the Coley-Jacobson 
\cite{CJ} type and have been used in \cite{trans_planck} to study the
trans-Planckian problem for cosmological perturbations in the context of toy
models.
 
As is obvious from the above equations, in the "healthy extension" of HL gravity 
we have to consider two effects in the evolution of fluctuations in the UV
regime which are different from what is obtained using GR: firstly, the UV 
modification of the dispersion relation introduced by the UV complete action leads
to a different form of the initial state and to a different evolution of the
modes on trans-Planckian scales, and secondly there is an extra degree of 
freedom for scalar metric fluctuations which is active in the UV. Through
the mode coupling, fluctuations in this second mode induce growing perturbations
of the curvature fluctuations. If we imagine
imposing initial conditions for all fluctuation modes at some initial time $t_i$,
then short wavelengths are subject to the presence of the source term for
a longer period than long wavelengths. Hence, one might expect that mode
mixing will lead to important deviations from scale-invariance \footnote{We 
remind the reader that in the toy models discussed in \cite{trans_planck} with
dispersion relations leading to non-adiabatic mode evolution is was precisely
the fact that short wavelength spend a longer period of time on sub-Planck
wavelengths that led to the blue tilt of the spectrum of curvature perturbations
obtained.}. 

The effects of modified dispersion relations on the spectrum of fluctuations in
inflationary cosmology was studied in detail in \cite{trans_planck} (see
also \cite{TPrev} for a review). It was found that if the dispersion
relation is of the type for which the adiabatic evolution of the mode function
is maintained, then no deviations from scale invariance of the perturbation
spectrum results. The dispersion relation which we obtain is in this class. 
Hence, we expect that deviations from scale-invariance of the spectrum of
curvature fluctuations in an inflationary universe in HL gravity can only
come from the mode mixing effect.

\begin{figure}[H]
\centering
\includegraphics[width=6cm,height=6cm,scale=0.6]{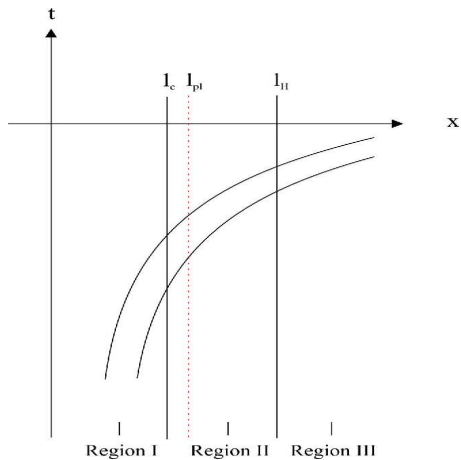} 
\caption{Space time diagram representing the perturbations modes in the UV regions: 
Region I, where the $k^{6}$ term dominates and  Region II, where $w_{k} \propto k$; 
and the IR Region III. In red we see the Planck length showing that the Region I 
must be sub-Planck.}
\label{regions}
\end{figure}

After these preliminary discussions, we turn to the solution of the
above system of coupled equations. In the linear approximation which
we are using each Fourier mode evolves independently. The behaviour
of the mode functions is different in the three regions of space-time
indicated in Figure \ref{regions}. Region I is the region where the
$k^6$ term in the dispersion relation dominates, i.e. where 
$\lambda \ll l_{c}$. The wavelength $l_c$ where the $k^6$ term
ceases to dominate is given by the mass parameters in the higher
space derivative terms in the HL action. Based on cosmic ray constraints
\cite{guy} we know that $l_{c}$ must be smaller than the Planck length
$l_{pl}$. In principle there could be a small region of wavelengths where
the $k^4$ term in the dispersion relation dominates. However, this would
require tuning of parameters. We will assume that we
can neglect the term in the dispersion relation proportional to  $k^{4}$, and
this is justified if the conditions $\alpha ^{2} \geq l_{0}^{2} \beta ^{2}$
(for  $u_{k}$) and $\delta ^{2} \geq l_{0}^{2} \epsilon ^{2}$ (for  $v_{k}$) 
are satisfied. We will hence assume that for 
$\lambda \gg l_{c}$ the linear term in the dispersion relation dominates.
Thus there is a region (Region II with $l_{c}\ll \lambda \ll l_{H}$, $l_{H}$
being the Hubble radius) where the dispersion relation is linear, as in 
standard GR theory. However, since the UV/IR scale is given by the Hubble 
radius, $l_{H}$, in Region II the extra degree of freedom is still dynamical 
and the interaction between the degrees of freedom must be considered. 
Region III is the IR domain where wavelengths are larger than
the Hubble radius. In Region III where $\lambda \gg l_{H}$ the 
extra degree of freedom disappears. The perturbations will freeze out and 
$\omega_{k,III}^{2} = a^{''}/a$. Hence, in this region the evolution of 
curvature fluctuations is as in GR. We wish to calculate the power spectrum 
at the time $\eta _{2}(k)$ where the perturbation mode $k$ becomes larger than 
$l_{H}$. 
 
\vspace{0.5cm}
 
To construct the general solution to this system, we need to match the different 
solutions from different regions at the transition times. Considering an expansion 
of the universe of power-law inflation type, $a(\eta )=l_{0} \eta ^{(1-b)/2}$
(note that we recover de Sitter inflation when $b=3$), 
the matching must be performed at the times:
\begin{eqnarray}
|\eta _{1}^{u}|^{(1-b)/2}=\left( \frac{\alpha}{c_{s}}\right)^{1/2} \frac{k}{l_{0}}\,,\hspace{2cm} |\eta _{1}^{v}|^{(1-b)/2}=\left( \frac{\delta}{\tilde{c}_{s}}\right)^{1/2} \frac{k}{l_{0}}\,,
\label{eta_1}
\end{eqnarray}
which is the transition between Regions I and II for the two modes, with 
$l_{c}^{u}=(2\pi )(\alpha/c_{s})^{1/2}$ and 
$l_{c}^{v}=(2\pi )(\delta/\tilde{c}_{s})^{1/2}$. 
Considering the definition of the coefficients (\ref{coef_u_v}), we can see that: 
\begin{align}
l_{c}=\left( \frac{4 \tilde{V}_{6}}{\frac{\tilde{V}_{4}}{2\eta _{4}}-2V_{1}} \right) ^{1/4} l_{pl}= C l_{pl}\,, \qquad \tilde{l}_{c}=\left( 8g_{7}-3g_{8}+2\frac{\eta _{5}^{2}}{\eta _{4}}  \right)^{1/4} l_{pl}=\tilde{C} l_{pl}\,.
\end{align}
From the constraints imposed by Gamma-Rays \cite{guy} $l_{c}$ and $\tilde{l}_{c}$ 
cannot be larger than the Planck length. We do not want them to be much smaller,
either, otherwise the effective field theory description of HL gravity
becomes questionable. Hence, we assume that the two lengths are a bit smaller
but close to the Planck length, and hence $C$ and $\tilde{C}$ are of order one. 

The transition between Regions II and III takes place at the times
\begin{eqnarray}
|\eta _{2}^{u}|=\frac{\sqrt{b^{2}-1} }{2}\frac{1}{c_{s}k}\,,\hspace{2cm} |\eta _{2}^{v}|=\frac{\sqrt{b^{2}-1} }{2}\frac{1}{c_{s}k}\,,
\label{eta_2}
\end{eqnarray}
with $l_{H}^{u}=4\pi c_{s}l_{0}|\eta|^{(3-b)/2}/\sqrt{b^{2}-1}$ and 
$l_{H}^{v}=4\pi \tilde{c}_{s}l_{0}|\eta|^{(3-b)/2}/\sqrt{b^{2}-1}$.
 
\subsubsection{Initial Conditions}

Before solving this system of equation we need to determine the initial conditions. 
Since in inflationary cosmology initial classical fluctuations redshift, it is
usually assumed that the initial conditions are given by quantum vacuum
perturbations, and that the system that will evolve following the classical equations
afterwards. In order to determine this initial state we will make use of the 
Hamiltonian formalism and canonically quantize the system. Since the
Hamiltonian in HL gravity is different from that in GR on the UV scales which
are relevant to us, the vacuum state (defined as the state which minimizes
the Hamiltonian \cite{trans_planck}) will differ.

From (\ref{S2_conforme}), it follows that the conjugate momenta for
the fluctuation modes are:
\begin{eqnarray}
p_{u_{k}}=\frac{\partial \delta \mathcal{L}_{k}}{\partial u_{k}^{'}}=2u_{k}^{'}+\frac{F_{\phi _{3}\phi _{4}}a}{\sqrt{\lambda _{1} \lambda _{2}}} v_{k}\,,\,\,\,\,\,\,\,\,\,\,p_{v{k}}=\frac{\partial \delta \mathcal{L}_{k}}{\partial v_{k}^{'}}=2v_{k}^{'}+\frac{\tilde{F}_{\phi _{3}\phi _{4}}a}{\sqrt{\lambda _{1} \lambda _{2}}} u_{k}\,.
\end{eqnarray}
Hence we can write the Hamiltonian of the system for each mode:
\begin{eqnarray}
\mathcal{H}_{k}=p_{u_{k}}u^{'}_{k}+p_{v_{k}}v^{'}_{k}-\delta \mathcal{L}_{k}=(u_{k}^{'2}-\omega _{u}^{2} u_{k}^{'})+(v_{k}^{'2}-\omega _{v}^{2} v_{k}^{'})+M^{2}_{uv} u_{k}v_{k}\,.
\end{eqnarray}

The canonical quantization can now be done. We promote the variables to operators 
that satisfy the canonical commutation relations:
\begin{alignat}{1}
&\left[\hat{u}_{k} (\eta),\hat{u}_{k^{'}} (\eta) \right]=\left[\hat{p}_{u_{k}} (\eta),\hat{p}_{u{k^{'}}} (\eta) \right]=0\,, \\
&\left[\hat{u}_{k} (\eta),\hat{p}_{u{k^{'}}} (\eta) \right]=\delta ^{(3)}(k-k^{'})\,,\label{comuta_uv}
\end{alignat}
that come from $[\phi (x,t),p_{\phi}(y,t)]=i\delta ^{(3)} (x-y)$ (all the others
are equal to zero). We have the same relations for $v_{k}$. 

As the Region I is sub-Planck, the initial conditions $t_{i}$ are given in the 
deep UV. In this limit, the term that dominates is the one proportional to
$k^{6}$ and the expansion of the universe can be neglected. The interaction term 
is also neglected initially and our system can be considered as a free system 
in a Minkowsi space-time. With this $p_{u_{k}}=2u_{k}^{'}$ and $p_{v_{k}}=2v_{k}^{'}$.

The fields of the non-interacting system can be expanded in
terms of creation and annihilation operators in the standard way:
\begin{alignat}{1}
u(x)&=\sqrt{\hbar}\int \frac{d^{3}p}{(2\pi)^{3/2}}\frac{1}{\sqrt{2\omega ^{I}_{u}}} \left( \hat{a}_{k}e^{ikx}+\hat{a}_{k}^{\dag}e^{-ikx}\right) \,,\\
v(x)&=\sqrt{\hbar}\int \frac{d^{3}p}{(2\pi)^{3/2}}\frac{1}{\sqrt{2\omega ^{II}_{v}}} \left( \hat{b}_{k}e^{ikx}+\hat{b}_{k}^{\dag}e^{-ikx}\right) \,,
\end{alignat}
where $\omega ^{I}_{u}=k^{6} \alpha ^{2}/a^{4}$ and 
$\omega ^{II}_{v}=k^{6} \delta ^{2}/a^{4}$. We can write the Fourier components of 
the fields in terms of the annihilation an creation operators:
\begin{alignat}{1}
&u_{k}=\sqrt{\frac{\hbar}{2\omega ^{I}_{u}}}\left( \hat{a}_{k}+\hat{a}_{-k}^{\dag } \right)\,, \,\,\,\,\,\,\,p_{u{k}}=-i\sqrt{\frac{\hbar \omega ^{I}_{u}}{2}}\left( \hat{a}_{k}-\hat{a}_{-k}^{\dag } \right)\,,\\
&v_{k}=\sqrt{\frac{\hbar}{2\omega ^{II}_{v}}}\left( \hat{b}_{k}+\hat{b}_{-k}^{\dag } \right)\,, \,\,\,\,\,\,\,p_{v{k}}=-i\sqrt{\frac{\hbar \omega ^{II}_{v}}{2}}\left( \hat{b}_{k}-\hat{b}_{-k}^{\dag } \right)\,.
\end{alignat}
The creation and annihilation operators obey the commutation relations 
$[\hat{a}_{k},\hat{a}_{k^{'}}^{\dag }]=\delta ^{(3)} (k-k^{'})$ and 
$[\hat{a}_{k},\hat{a}_{k^{'}}^{\dag }]=\delta ^{(3)} (k-k^{'})$, with all the 
others equal to zero. These commutation relations are valid if the 
following normalization condition holds:
\begin{equation}
u_{\mathbf{k}}^{'}u_{\mathbf{k}}^{*}-u_{\mathbf{k}}^{*'}u_{\mathbf{k}}=2i\,,\,\,\,\,\,\,\,\,\,\,\,v_{\mathbf{k}}^{'}v_{\mathbf{k}}^{*}-v_{\mathbf{k}}^{*'}v_{\mathbf{k}}=2i\label{eq:normalizacao}
\end{equation}
which is the Wronskian of the classical solutions. This normalization 
allows us to fix the amplitude of $u_{\mathbf{k}}\left(\eta\right)$ and 
$v_{\mathbf{k}}\left(\eta\right)$
such that it is compatible with the Heisenberg uncertainty principle.

We can now determine the initial conditions for our modes. Because of the 
normalization conditions (\ref{eq:normalizacao}), this task is reduced to the 
determination of the effective frequency at the initial time. As
mentioned before, we choose our initial state to be the one that
minimizes the energy. This is a good definition for the initial vacuum state 
since it generalizes how the vacuum state is usually defined to the case
of a modified dispersion relation. As these fluctuations are initially oscillating
like they would in a Minkowski space-time, our initial conditions prescription
can be viewed as an application of the Einstein Equivalence Principle
to our problem. Making the ansatz
\begin{equation}
u_{\mathbf{k}}=r_{\mathbf{k}}(\eta )\mathrm{e}^{i\gamma_{\mathbf{k}}}\,,\,\,\,\,\,\,\,\,\,\,v_{\mathbf{k}}=\tilde{r}_{\mathbf{k}}(\eta )\mathrm{e}^{i\tilde{\gamma_{\mathbf{k}}}}\,.
\end{equation}
we see that in order for these variables in this form to obey the commutation 
relations (\ref{comuta_uv}) and the Wronskian condition (\ref{eq:normalizacao}), 
we have that $r_{k}^{2} \gamma _{k}^{'}=1$ and 
$\tilde{r}_{k}^{2} \tilde{\gamma } _{k}^{'}=1$.

The energy of the system is:
\begin{alignat}{1}
E_{\mathbf{{k}}} & =\frac{1}{2}\left(\left| u_{\mathbf{k}}^{'}\right|^{2}+\omega ^{I\,2}_{u}\left| u_{\mathbf{k}}\right|^{2}\right)+\frac{1}{2}\left(\left| v_{\mathbf{k}}^{'}\right|^{2}+\omega _{v}^{II\,2}\left| v_{\mathbf{k}}\right|^{2}\right)\\
 & =\frac{1}{2}\left[\left(r_{\mathbf{k}}^{'}\right)^{2}+\frac{1}{r_{\mathbf{k}}^{2}}+\omega _{u}^{I\,2} r_{\mathbf{k}}^{2}\right]+\frac{1}{2}\left[\left( \tilde{r}_{\mathbf{k}}^{'}\right)^{2}+\frac{1}{\tilde{r}_{\mathbf{k}}^{2}}+\omega _{v}^{II\,2}\tilde{r}_{\mathbf{k}}^{2}\right]\,.
\end{alignat}
Considering the minimal quantum fluctuations allowed by the uncertainty principle, 
the minimum energy is $E_{u,k}\left(\eta_{i}\right)=\omega _{u}^{I}$ and 
$E_{v,k}\left(\eta_{i}\right)=\omega _{v}^{II}$ which corresponds to:
\begin{alignat}{1}
r_{\mathbf{k}}\left(\eta_{i}\right)=\omega_{u}^{I\,-1/2}\,&, \qquad \tilde{r}_{\mathbf{k}}\left(\eta_{i}\right)=\omega_{v}^{-1/2}\,, \\
r_{\mathbf{k}}^{'}\left(\eta_{i}\right)=0\, &,\qquad
\tilde{r}_{\mathbf{k}}^{'}\left(\eta_{i}\right)=0\,.
\end{alignat}
With this, the initial conditions are:
\begin{alignat}{1}
u_{k}(\eta _{i})=\frac{1}{\sqrt{\omega ^{I}_{u}(\eta _{i})}}=\frac{a(\eta _{i})}{\alpha ^{1/2} k^{3/2}}\, &,\qquad v_{k} (\eta _{i})=\frac{1}{\sqrt{\omega ^{II}_{v}(\eta _{i})}}=\frac{a(\eta _{i})}{\delta ^{1/2} k^{3/2}}\,,\\
u^{'}_{k}=i\sqrt{\omega _{u}^{I}}=i\frac{\alpha ^{1/2}k^{3/2}}{a(\eta _{i})}\,&, \qquad v^{'}_{k}=i\sqrt{\omega _{v}^{II}}=i\frac{\delta ^{1/2}k^{3/2}}{a(\eta _{i})}\,. \label{cond_ini}
\end{alignat}

\subsubsection{Solving the Equation of Motion}

Having determined the initial conditions we can now solve the equations of 
the system (\ref{eom_uv}) in each region and then match them. The only degree 
of freedom that is propagating in the IR is $u_{k}$ and this is the observable 
one that we are interested in obtaining the power spectrum in order to 
compare with the results obtained in standard inflation. Because of this, we are 
only interested in calculating the solution of $u_{k}$. The perturbation 
$v_{k}$ acts as an "entropy" perturbation, analogous to what happens in
multi-field inflation models \cite{multifield}, and will source the 
"adiabatic" perturbation $u_{k}$ in Regions I and II, decoupling in Region III.

The equations of motion that we want to solve are:
\begin{eqnarray}
u_{k}^{''}+\left( \frac{\alpha ^{2}k^{6}}{a^{4}}+c_{s}^{2}k^{2}-\frac{a^{''}}{a} \right) u_{k}=S_{u} \,, \\
v_{k}^{''}+\left( \frac{\delta ^{2}k^{6}}{a^{4}}+\tilde{c}_{s}^{2}k^{2}-\frac{a^{''}}{a} \right) v_{k}=S_{v} \,. \label{eom_uv2}
\end{eqnarray}
where 
\begin{eqnarray}
S_{u} = -v_{k}^{'} a \Xi _{2} + 
v_{k} \left( \frac{\gamma ^{2}k^{4}}{a^{2}}+\Xi _{3} a^{2} \right) \,, \qquad
S_{v} = -u_{k}^{'} a \Omega _{2} +
u_{k} \left( \frac{\zeta ^{2}k^{4}}{a^{2}}+\Omega _{3} a^{2} \right) \,,
\end{eqnarray} 
are the interaction terms. 

The homogeneous equations of motion have the same form of the equations of motion 
studied in \cite{trans_planck} in the case of the Corley-Jacobson dispersion 
relation. In that work it was found that the evolution depends sensitively on
the signs of the coefficients of the terms in the dispersion relation. In our case 
we know from the construction of the "healthy" HL action that the coefficients
$\alpha$, $\delta$, $c_{s}$ and $\tilde{c}_{s}$ must be positive to avoid exponential,
\textit{ghost} and tachyonic instabilities. If we evaluate the "adiabaticity coefficient"  for our dispersion relation
in the UV region and in the case of de Sitter inflation, we obtain
\begin{eqnarray}
Q (k,\eta ) \equiv \left| \frac{ \alpha ^{2} k^{4} \eta ^{2b} + c_{s}^{2} l_{0}^{4} \eta ^{2}}{\alpha ^{2} k^{4} \eta ^{2b} - c_{s}^{2} l_{0}^{4} \eta ^{2}} \right| \,.
\end{eqnarray}
and it is hence clear that $Q(k, \eta) > 1$ which implies that the adiabaticity 
condition is satisfied. Hence, we know that the 
state tracks the instantaneous vacuum state, and thus the homogeneous
solutions approach those of GR for $\lambda > l_{pl}$. The homogeneous
evolution alone does not lead to differences in the power spectrum of
cosmological perturbations compared to what is obtained in standard
inflation.

On the other hand, in both Regions I and II these equations are going to have 
both a homogeneous and a particular solution, because these are the UV regions 
where the extra degree of freedom is dynamical and mode mixing occurs. 

Let us now turn to the effect of mode interactions in the two UV regions,
and proceed to evaluate the analytic solutions of the equations of motion 
in the presence of the interaction in each region. For simplicity, we will 
choose $\alpha =\delta$ and $c_{s}=\tilde{c}_{s}$, in order to only have a
common value of $l_{c}$ and ${l_{H}}$ for both modes.

\vspace{0.5cm}

In Region I, the equation of motion reduces to:
\begin{eqnarray}
u_{I\,k}^{''}+ \frac{\alpha ^{2}k^{6}}{a^{4}} u_{I\,k}=S_{u} \,, \\
v_{I\,k}^{''}+ \frac{\delta ^{2}k^{6}}{a^{4}} v_{I\,k}=S_{v} \,. \label{eom_reg_1}
\end{eqnarray}
We will only show here the calculations for  $u_{I\,k}$, since the solution for 
$v_{I\,k}$ is identical, only changing the coefficients. 

We are going to solve this equation using the Green's function method with 
Born approximation. This means that our solution is going to be the general 
solution of the homogeneous equation $Lu=0$, where $L=(d^{2}/d\eta ^{2})+\omega ^{2}$, 
plus a particular solution of the inhomogeneous equation. The particular solution 
is constructed using the fundamental solution of $LG(\eta , a)=\delta (\eta - a)$, 
where $G(\eta, a)$ is the Green's function. The solution is:
\begin{equation}
u_{k}(\eta )=u_{k}^{h}(\eta)+\int _{-\infty}^{+\infty}f_{u}(a)G_{u}(\eta ,a)da
\end{equation}
where $u^{h}$ is the homogeneous one that obeys the initial conditions 
(\ref{cond_ini}), and second term on the right hand side is the particular 
solution constructed from the Green's function.

The homogeneous solution has two independent solutions:
\begin{equation}
u_{I\,k}^{h}(\eta)=\underbrace{A_{1}|\eta |^{1/2}J_{1/2b}(z)}_{u_{1,k}}+\underbrace{A_{2}|\eta |^{1/2}J_{-1/2b}(z)}_{u_{2,k}}\,,\label{sol_1}
\end{equation}
where $z=g|\eta |^{b}/b$ e $g=\alpha k^{3}/l_{0}^{2}$. For the
$v_{I\,k}$ equation we would have $\tilde{z}=\tilde{g}|\eta |^{b}/b$ 
where  $\tilde{g}=\delta k^{3}/l_{0}^{2}$. To determine  $A_{1}$ and 
$A_{2}$ we use the initial conditions discussed earlier. Using the 
Wronskian of the Bessel functions,  
$[J_{-\nu}J_{\nu -1}+J_{\nu +1}J_{\nu}](z)=2\sin [\pi /(2b)]/\pi z$, we obtain
\begin{alignat}{1}
A_{1}=\frac{g\pi}{2b\sin (\pi/(2b))}|\eta _{i}|^{b-1/2} u_{I\,k}(\eta _{i})J_{1-\frac{1}{2b}}(z_{i})\left[1-\frac{|\eta _{i}|^{1-b}}{g}\frac{u_{I\,k}(\eta_{i})}{u^{'}_{I\,k}(\eta_{i})}\frac{J_{-1/2b}(z_{i})}{J_{1-\frac{1}{2b}}(z_{i})} \right]\,,\\
A_{2}=\frac{g\pi}{2b\sin (\pi/(2b))}|\eta _{i}|^{b-1/2}u_{I\,k}(\eta _{i})J_{\frac{1}{2b}-1}(z_{i})\left[1+\frac{|\eta _{i}|^{1-b}}{g}\frac{u_{I\,k}(\eta_{i})}{u^{'}_{I\,k}(\eta_{i})}\frac{J_{1/2b}(z_{i})}{J_{\frac{1}{2b}-1}(z_{i})} \right]\,,
\end{alignat}
unless $1/2b$ an integer, which we assume is not the case.

In this regime, $z_{i}$ is large and proportional to 
$l_{H}/\lambda (\eta _{i}) \gg 1$. Thus, we can use the asymptotic expansion 
of the Bessel functions for large arguments, 
$J_{\nu}(x)=(2/\pi x)^{1/2} \cos (x-\nu \pi/2-\pi/4)$, and rewrite the 
coefficients as:
\begin{eqnarray}
A_{1} \approx \mp i \left(\frac{\pi g}{2b} \right)^{1/2}|\eta _{i}|^{(b-1)/2} \frac{u_{I\,k}(\eta _{i})}{\sin (\pi /2b)}e^{\pm ix_{i}}\,,\qquad
A_{2} \approx \pm i \left(\frac{\pi g}{2b} \right)^{1/2}|\eta _{i}|^{(b-1)/2} \frac{u_{I\,k}(\eta _{i})}{\sin (\pi /2b)}e^{\pm iy_{i}}\,,\label{coef_region_1}
\end{eqnarray}
where $x_{i}(\eta )\equiv z_{i}(\eta )+(\pi/4b)-(\pi/4)$ and $y_{i}(\eta )\equiv z_{i}(\eta)-(\pi/4b)-(\pi/4)$.

We can rewrite the particular solution (in the case where we have two independent solutions 
for the homogeneous equation (\ref{eom_reg_1})) as:
\begin{equation}
\delta u (\eta )=u_{1,\,k}(\eta )\int _{\eta _{i}}^{\eta _{1}(k)}d\eta ^{'} W^{-1}_{u}u_{2,\,k}(\eta ^{'})f_{u}(\eta ^{'})-u_{2,\,k}(\eta )\int _{\eta _{i}}^{\eta _{1}(k)}d\eta ^{'} W^{-1}_{u}u_{1,\,k}(\eta ^{'})f_{u}(\eta ^{'})\,,
\end{equation}
where \textit{W} is the Wronskian of the solutions with index $1$ and $2$ which in 
Region I is $W=-2A_{1}A_{2}b\sin (\pi/2b)/\pi $. The integration limits in Region I 
are from $\eta _{i}$ until the cutoff $\eta _{1}^{v}(k)$, which is the transition 
between Regions I and II. To write the interaction, $S_{u}$ we use the Born 
approximation where only the homogeneous solution of $v_{k}$ is considered. 
The solution for $v_{I,k}^{h}$ is the same as the one for $u_{I,k}^{h}$ given 
in (\ref{sol_1}).

We can now evaluate the particular solution. This is done in Appendix 4, where 
we considered the general case and the case of de Sitter inflation. Here we
only discuss the second  case, because in this limit we can evaluate the integrals
exactly. 

As we can see from (\ref{gamma_1_b3}) and (\ref{gamma_2_b3}) that the integrals 
only depend on the combination $k\eta$ and that the integration limit 
$k\eta _{1}=const.$ This shows the integrands for different modes are
related by time translation, as is the late time cutoff. The cutoff
corresponding to the transition between Regions I and II also obeys
this symmetry. The only source of different evolution between different
modes is the initial time which is fixed for all modes \footnote{This
is the usual feature in inflationary cosmology that the only source
of violation of the time translation symmetry between the evolution
of different modes is the initial condition at a fixed time.}.
If the integrals are dominated by earliest times, we would expect a
the mode mixing to lead to a large deviation from scale-invariance,
whereas if the integrals are dominated by the contribution close to
the transition point between Regions I and II then scale-invariance
will be preserved. We now have to check if the dependency on $k$ which
arises from the  initial condition is strong enough to affect the power 
spectrum in an important way. 

The terms $1/(k\eta _{i})^{n}$ in (\ref{sol_part_I}), where $n=3$, $2$ and $6$, 
respectively, are always very small since they depend on 
$\left( \lambda (\eta _{i})/l_{c} \right)^{n}$ which is small since the initial 
length scale is much smaller than $l_{c}$. Therefore these terms can be 
neglected and the solution is given by:
\begin{align}
\delta u_{I} \left( \eta _{1} \right) & =\mp\frac{1}{c_{s}^{3/2}k^{1/2}} \left\{ -\frac{\Xi_{2}}{8\pi ^{2}} l_{c} e^{-2iz_{1}}\left(\cos\left(z_{1}-\frac{\pi}{3}\right)c_{1}^{-}+\cos\left(z_{1}-\frac{\pi}{6}\right)c_{2}^{-}\right)+\right. \nonumber \\
 & \left.+\left[\frac{3\pi \gamma^{2}}{c_{s}}\frac{1}{l_{c}^{2}}e^{-2iz_{1}} -\frac{3\sqrt{2}}{16\pi ^{2}}\Xi_{3} l_{c}^{2} e^{-2iz_{1}}\right]\left(\cos\left(z_{1}-\frac{\pi}{3}\right)c_{1}^{+}+\cos\left(z_{1}-\frac{\pi}{6}\right)c_{2}^{+}\right) \right\} \,,
 \label{up_eta1}
\end{align}
where $c_{1}^{\pm}=e^{\pm ix_{i}} \pm (-1)^{5/6}e^{\pm i y_{i}+i5\pi /3}$ and 
$c_{1}^{\pm}=e^{\pm ix_{i}+i7\pi /3}  (-1)^{7/6} \pm e^{\pm i y_{i}}$ are  
oscillating terms in $k$. We can read off the important result that the 
solution maintains the overall scale invariance of the power spectrum since
the overall amplitude is proportional to $k^{-1/2}$. The phase $z_1$ is
also independent of scale. However, we have an oscillatory dependence on
$e^{\pm i z_{i}}$ (since $z_{i}=z_{i}(k)$) coming from the intial integration
limit (the initial time). Note that the oscillations have a frequency
which is proportional to $k^3$. 

Note that the overall amplitude of the inhomogeneous term is smaller than
that of the homogeneous one, as one can check by inserting the expressions
for the various constants. 

\vspace{1cm}

We now move on to a study of the solutions in Region 2
in which the equation of motion has a linear dispersion relation:
\begin{equation}
u_{II\,k}^{''}+c_{s}^{2}k^{2}u_{II\,k}=S_{u}\,,
\end{equation}
with a homogeneous solution given as a linear combination of
two basis solutions
\begin{equation}
u_{II\,k}(\eta )=B_{1}e^{ic_{s}k\eta }+B_{2}e^{-ic_{s}k\eta } \, .
\end{equation}

As our homogeneous solution we take the one which matches with the
full solution at the end of Region I (the effect of the mode mixing
in Region II will yield the particular solution in Region II).
We need to match the solutions $u_{I\,k}$ and $u_{II\,k}$ when the transition between 
the regions occur at $\eta _{1}$, to determine the coefficients $B_{1}$ and $B_{2}$. 
This gives us: 
\begin{alignat}{1}
B_{1}=\mp \frac{u_{I\,k}(\eta _{i})e^{-ic_{s} k\eta _{1}}}{2\sin (\pi /2b)} |\frac{\eta _{1}}{\eta _{i}}|^{(1-b)/2}e^{\pm ix_{i}} \left(e^{-iy_{1}}-e^{\mp (i \pi /2b)}e^{-i x_{1}} \right)+\frac{e^{-ic_{s}k\eta _{1}}}{2}\left( \delta u_{I}(\eta _{1})-\frac{i}{c_{s}k}\delta u_{I}^{'} (\eta _{1}) \right)\,,\\
B_{2}=\mp \frac{u_{I\,k}(\eta _{i})e^{ic_{s}k\eta _{1}}}{2\sin (\pi /2b)} |\frac{\eta _{1}}{\eta _{i}}|^{(1-b)/2}e^{\pm ix_{i}} \left(e^{i y_{1}}-e^{\mp (i \pi /2b)}e^{ix_{1}} \right)+\frac{e^{ic_{s}k\eta _{1}}}{2}\left( \delta u_{I}(\eta _{1})-\frac{i}{c_{s}k}\delta u_{I}^{'} (\eta _{1}) \right)\,.
\label{B}
\end{alignat}

To determine the particular solution in this region, we use the same procedure as 
in Region I. The particular solution is calculated in Appendix 4. Again, considering 
the case of de Sitter inflation, it follows from (\ref{delta_u_II_b3}) that both
of the integration limits are constant, $k\eta _{1}=(c_{s}/\alpha)^{1/2} l_{0}=const.$ 
and $k\eta _{2}=\sqrt{2}/c_{s}=const.$.The integration range will hence be independent 
of $k$.  On the other hand, a dependence on
$k$ enters via the homogeneous solution in the source term. The k-dependence of 
the particular solution comes completely from the coefficients $\tilde{B}_{1}$ and 
$\tilde{B}_{2}$ (the coefficients analogous to the above $B_1$ and $B_2$ which
appear in the homogeneous solution for $v_k$) which are both proportional to $k^{-1/2}$, indicating
that the overall amplitude of the inhomogeneous term will respect scale invariance. 
However, like in the contribution to the inhomogeneous term in Region I, we still 
have oscillation.

Evaluating the 9 integrals present in the solution (\ref{delta_u_II_b3}), 
and using the asymptotic expansion of the exponential integral function for 
large arguments, $Ei(z) \sim e^{-z}/z+...$ \cite[pg. 231]{ abramowitz} we 
consider the dominant contribution from each term. Considering that the 
contributions of the particular solution in Region I and its derivative in the 
coefficients $\tilde{B}_{1}$ and $\tilde{B}_{2}$, given in (\ref{B}), are 
smaller that of the homogeneous one, we can neglect them. With this approximation, 
the particular solution becomes:
\begin{equation} 
\delta u_{II} (\eta _{2})=\pm \frac{i}{c_{s}^{3/2}k^{1/2}}e^{\pm ix_{i}}\left[ -i\frac{\sqrt{2}}{2}\Xi _{2} l_{H} \ln \left( \frac{1}{2\pi ^{2}} \frac{l_{H}}{l_{c}} \right) \alpha +\frac{\pi ^{2}}{2}\frac{\gamma ^{2}}{l_{c}^{2}} \lambda+\frac{\sqrt{2}}{4\pi ^{2}} \Xi _{3} l_{H}^{2} \left( \tilde{\alpha} \sin (2\sqrt{2})-\alpha \right) \right]\,,
\label{final}
\end{equation}
where 
\begin{eqnarray}
\alpha &=& \sin (\sqrt{2}(l_{H}/l_{c})+y_{1})-\sin (\sqrt{2}(l_{H}/l_{c})+x_{1}\pm\pi/6)\,, \\ 
\tilde{\alpha} &=& \cos (\sqrt{2}(l_{H}/l_{c})+y_{1})
-\cos (\sqrt{2}(l_{H}/l_{c})+x_{1}\pm\pi/6) \,, \nonumber \\
\lambda &=& \sin \left[ \sqrt{2} \left( 1+3l_{H}/l_{c} \right)+y_{1} \right] 
-\sin \left[ \sqrt{2} \left( 1+3l_{H}/l_{c} \right)+x_{1} \pm \pi/6 \right]\,, \nonumber
\end{eqnarray}
are constants containing terms that depend on  $z_{1}$ which is independent of $k$.

Comparing term by term, this particular solution dominates over the one from 
Region I.  The particular solution is oscillatory in $k$ (with a frequency
which is proportional to $k^3$) via the overall coefficient $e^{\pm ix_{i}}$.
Comparing the amplitude of the particular solution with that of the
homogeneous one, we see that the overall amplitudes (outside the square
bracket of (\ref{final})) are the same. The relative amplitudes of the
oscillatory and constant terms in the mode functions is given by the
coefficients inside the square bracket of (\ref{final}). Taking
the new scale appearing in the HL Lagrangian to be comparable to the
Planck scale we find that the amplitudes of the first two terms are
of the order one, not suppressed by factors of $H / m_{pl}$ as the
oscillations obtained in \cite{hypersurface} are. The
amplitude of the third term is suppressed by the inflationary slow-roll
parameter.

Since
\begin{equation}
z_{i} = \frac{\sqrt{2}}{3} \left( \frac{l_{c}}{\lambda(\eta_{i})} \right)^2 
\left( \frac{l_{H}}{\lambda(\eta_{i})} \right)\,,
\label{z_i}
\end{equation}
we see that the frequency of oscillation is set by the initial time
and is very large in units of $\frac{l_H}{\lambda(\eta_i)}$ (since
the second factor on the right hand side of (\ref{z_i})
is much larger than unity). Hence, these oscillations will be
very hard to detect observationally given a finite frequency
resolution of an experiment.

To evaluate the power spectrum of curvature fluctuations we need
to have the solution in Region III and match it with the full solution in 
Region II at $l_{H}$ crossing.

In Region III, we have the equation:
\begin{equation}
u_{III\,k}^{''}+\frac{a^{''}}{a}u_{III\,k}=0\,
\end{equation}
with solution $u_{III\,k}=C\,a(\eta )$. Now, we need to match the solutions when the 
mode exits the Hubble radius at $\eta _{2}$ to determine $C$. At $\eta _{2}$ the 
solution in Region II is given by:
\begin{align}
u_{II\,k}(\eta _{2})&=\mp \frac{1}{c_{s}^{3/2}k^{1/2}}e^{\pm ix_{i}} \left( e^{ic_{s}(\bar{\eta}_{2}-\bar{\eta}_{1})} h+ e^{-ic_{s}(\bar{\eta}_{2}-\bar{\eta}_{1})} h^{*} \right) \nonumber \\
&+\frac{1}{2}\left( e^{ic_{s}(\bar{\eta}_{2}-\bar{\eta}_{1})} + e^{-ic_{s}(\bar{\eta}_{2}-\bar{\eta}_{1})}  \right) \left( \delta u_{I} (\eta _{1})-\frac{i}{c_{s}k} \delta u_{I}^{'} (\eta _{1}) \right)+\delta u_{II} (\eta _{2})\,,
\end{align}
where  $h=e^{-iy_{1}}-e^{-i(x_{1} \pm \pi /2)}$ and $h^{*}$ it complex conjugate. Thus,
the coefficient is $C=u_{II}(\eta _{2})/a$.

\subsubsection{Power Spectrum}

{F}inally we can evaluate the power spectrum of the $\phi _{3}$, the Mukhanov-Sasaki variable . 
This is the only propagating mode in the IR and will be the "adiabatic" perturbation produced 
during inflation.  The dimensionless power spectrum will be calculated at Hubble horizon crossing. 
Afterwards the perturbations freeze until re-enter the Hubble horizon later in the evolution of the universe. 
The power spectrum is given by:
\begin{align}
k^{3}P_{\phi}=k^{3}|C|^{2}&=\frac{k^{3}\eta _{2}^{2}}{ l_{0}^{2}} \left| u_{II}(\eta _{2}) \right| ^{2} \nonumber \\
&= \frac{4 \pi ^{2} k }{l_{H}^{2}} \left| \frac{1}{c_{s}^{3/2}k^{1/2}}e^{\pm ix_{i}} \left( e^{ic_{s}(\bar{\eta}_{2}-\bar{\eta}_{2})} h+ e^{-ic_{s}(\bar{\eta}_{2}-\bar{\eta}_{2})} h^{*} \right) \right.  \nonumber \\
&\left. +\frac{1}{2}\left( e^{ic_{s}(\bar{\eta}_{2}-\bar{\eta}_{2})} + e^{-ic_{s}(\bar{\eta}_{2}-\bar{\eta}_{2})}  \right) \left( \delta u_{I} (\eta _{1})-\frac{i}{c_{s}k} \delta u_{I}^{'} (\eta _{1}) \right)+\delta u_{II} (\eta _{2})  \right| ^{2} \,.
\end{align}
We can see from this solution that the amplitude of the power spectrum will be scale invariant since 
all terms are proportional to $k^{1/2}$, as we can see in (\ref{up_eta1}) and (\ref{final}). As 
already pointed out, we will have an oscillation in this power spectrum coming from terms that 
depend on $z_{i}$ and its mixing with the other terms in the sum. The relative amplitude
of the oscillations is of order one, show a characteristic $k$-dependent frequency, but are
too rapid to be detectable by an experiment with finite frequency resolution.

\section{Conclusion}

We have studied the generation and early evolution of cosmological fluctuations in the
``healthy extension" of Ho\v{r}ava-Lifshitz gravity, assuming that there is matter which
leads to an inflationary background cosmology. Since HL gravity can be viewed as
a consistent UV completion of gravity, it being
a power-counting renormalizable model for quantum gravity, inflation in the
context of HL gravity provides an interesting
background to study the {\it trans-Planckian problem} for inflationary cosmological
perturbations.

There are two reasons which could lead to large deviations from the usual predictions
of standard inflationary cosmology: firstly, the dispersion relation on trans-Planckian
scales is non-standard, and secondly there is a second scalar metric fluctuation mode
which acts as a source term in the evolution of the curvature fluctuations. The time
interval during which mode mixing occurs depends on the wavenumber $k$ of the
fluctuation mode.

The result of our analysis is that the overall scale-invariance of cosmological
perturbations is maintained in HL inflation. On the other hand, there are
oscillations in the amplitude of the spectrum as a function of $k$ whose relative
amplitude is of order one, and whose frequency shows a characteristic
scaling proportional to $k^3$, the power being determined by the power of
the highest power of the spatial derivative term in the HL action. The frequency
of oscillation is so high, however, that it is unlikely that an experiment with
finite frequency resolution will be able to detect these oscillations.

The reason why scale-invariance of the spectrum of cosmological perturbations
is maintained in HL inflation is that the dispersion relation for the fluctuation
modes satisfies the adiabaticity condition, and that the non-trivial initial
vacuum state evolves into the usual one once the wavelength drops to
sub-Planckian values. The oscillations in the spectrum come from the
particular solution of the mode evolution equation (induced by mode
mixing). Physically, the oscillations are a consequence of the fact that
setting initial conditions for all Fourier modes of the fluctuation variable
at a fixed initial time breaks the time-translation invariance of the
mode evolution.

\begin{acknowledgments}

The authors would like to thank Dr. Wang Yi for fruitful discussions. E.F. acknowledges financial
support from FAPESP (Funda\c c\~ao de Amparo \`a Pesquisa do Estado de S\~ao Paulo) and CNPq
(Conselho Nacional de Desenvolvimento Cient\'\i fico e Tecnol\' ogico). At McGill this research
was supported in part by an NSERC Discovery grant and by funds from the Canada Research
Chair program.

\end{acknowledgments}


\appendix

\section*{Appendix 1: Coefficients of the Second Order Action} 
\label{sec:appendix_1}

The coefficients of the second order Lagrangian of the Healthy extension of Horava-Lifshitz gravity
are
\begin{eqnarray}
d\left(\bar{k}\right)c_{\phi}&=&2\left(3\lambda-1\right)\frac{H^{2}}{\chi ^{2}}+\left(\lambda-1\right)f_{1}\left(\bar{k}\right)\frac{\bar{k}^{2}}{\chi ^{2}}\,, \hspace{1.5cm}
d\left(\bar{k}\right)c_{\psi}=2\left(3\lambda-1\right)\left[\frac{\dot{\phi}_{0}^{2}}{\chi ^{2}}+2f_{1}\left(\bar{k}\right)\bar{k}^{2}\right]\,,\\
d\left(\bar{k}\right)c_{\phi\psi}&=&4\left(3\lambda-1\right)H\frac{\dot{\phi}_{0}^{2}}{\chi ^{2}}\,.\hspace{1cm}
d\left(\bar{k}\right)f_{\phi}  =-\left(3\lambda-1\right)H\frac{\dot{\phi}_{0}^{2}}{\chi ^{4}}-\left(\lambda-1\right)\left[V_{0,\phi}+V_{4}\left(\phi_{0}\right)\bar{k}^{4}\right]\frac{\dot{\phi}_{0}}{\chi ^{4}}\,,\\
d\left(\bar{k}\right)f_{\psi}  &=&-24\left(3\lambda-1\right)H\Lambda+12\left(3\lambda-1\right)^{2}H^{3}-6\lambda\left(3\lambda-1\right)\frac{\dot{\phi}_{0}^{2}}{\chi ^{2}}H-12\left(3\lambda-1\right)\frac{V_{0}}{\chi ^{2}}\left(\phi_{0}\right)H\,,\\
d\left(\bar{k}\right)f_{\phi\psi}  &=&6\left(\lambda-1\right)\frac{\dot{\phi}_{0}^{2}}{\chi ^{2}}\Lambda-3\left(3\lambda-1\right)\left(3\lambda+1\right)\frac{\dot{\phi}_{0}^{2}}{\chi ^{2}}H^{2}-\frac{3}{2}\left(\lambda-1\right)\frac{\dot{\phi_{0}^{3}}}{\chi ^{4}}+\\
  &+&3\left(\lambda-1\right)\frac{V_{0}\left(\phi_{0}\right)}{\chi ^{2}}H-2\left(\lambda-1\right)\left[3f_{1}\left(\bar{k}\right)+2f_{2}\left(\bar{k}\right)\right]\frac{\dot{\phi}_{0}}{\chi ^{2}}\,\bar{k}^{2}\,,\\
d\left(\bar{k}\right)\bar{f}_{\phi\psi}  &=&4\left(3\lambda-1\right)\frac{V_{0,\phi}}{\chi ^{2}}\left(\phi_{0}\right)H-\left(3\lambda-1\right)\frac{\phi_{0}^{3}}{\chi ^{2}}-2\left(3\lambda-1\right)f_{1}\left(\bar{k}\right)\frac{\dot{\phi}_{0}\bar{k}^{2}}{\chi ^{2}}+4\left(3\lambda-1\right)\frac{V_{4}\left(\phi_{0}\right)H\bar{k}^{4}}{\chi ^{2}}\,.
\end{eqnarray}

\appendix

\section*{Appendix 2: Rotated Coefficients}
\label{sec:appendix_2}

The coefficients of the rotated second order Lagrangian are
\begin{align}
F_{\phi _{3}}&=\frac{1}{2} (1+\cos 2\theta ) f_{\phi} +\frac{1}{2} (1-\cos 2\theta ) f_{\psi}+\frac{\sin 2\theta}{2} (f_{\phi \psi } +\tilde{f}_{\phi \psi })\,. \\
F_{\phi _{4}}&=\frac{1}{2} (1-\cos 2\theta ) f_{\phi} +\frac{1}{2} (1+\cos 2\theta ) f_{\psi}-\frac{\sin 2\theta}{2} (f_{\phi \psi }+\tilde{f}_{\phi \psi })\,. \\
F_{\phi _{3} \phi _{4}}&=\frac{\sin 2\theta}{2} (f_{\psi }-f_{\phi})+\frac{1}{2} (1+\cos 2\theta ) f_{\phi \psi} -\frac{1}{2} (1-\cos 2\theta ) \tilde{f}_{\phi \psi}\,.\\
\tilde{F}_{\phi _{3} \phi _{4}}&=\frac{\sin 2\theta}{2} (f_{\psi }-f_{\phi})+\frac{1}{2} (1+\cos 2\theta ) \tilde{f}_{\phi \psi} +\frac{1}{2} (1-\cos 2\theta )f_{\phi \psi} \,.  \\
M^{2}_{\phi _{3}}&= \frac{1}{2} (1+\cos 2\theta ) m^{2}_{\phi} +\frac{1}{2} (1-\cos 2\theta ) m^{2}_{\psi}+\frac{\sin 2\theta}{2} m^{2}_{\phi \psi } \,. \\
M^{2}_{\phi _{4}}&= \frac{1}{2} (1-\cos 2\theta ) m^{2}_{\phi} +\frac{1}{2} (1+\cos 2\theta ) m^{2}_{\psi}-\frac{\sin 2\theta}{2} m^{2}_{\phi \psi } \,. \\
M^{2}_{\phi _{3} \phi _{4}}&=\frac{\sin 2\theta}{2} (m^{2}_{\psi }-m^{2}_{\phi})+\cos 2\theta \, m^{2}_{\phi \psi} \,. 
\end{align}
where:
\begin{eqnarray}
\sin 2\theta =\frac{c_{\phi \psi }}{\sqrt{(c_{\psi}-c_{\phi})^{2}+c^{2}_{\phi \psi}}}\,,\,\,\,\,\,\,\,\,\cos 2\theta =\frac{c_{\psi}-c_{\phi}}{\sqrt{(c_{\psi}-c_{\phi})^{2}+c^{2}_{\phi \psi}}}\,. \label{sin_cos} \, .
\end{eqnarray}
The general form of the $c_x$  coefficients it quite involved and will not be written
down here.  

\appendix

\section*{Appendix 3: Coefficients of the UV EOM} 
\label{sec:appendix_3}

The coefficients of the UV equation of motion are given by:
\begin{eqnarray}
\alpha ^{2}=4\frac{\tilde{V}_{6}}{\chi ^{4}}\,,\hspace{1cm}\beta ^{2}=\frac{2}{\chi ^{2}}(\tilde{V}_{2}+\tilde{V}_{4,\phi })\,, \hspace{1cm} c^{2}_{s}=\frac{\tilde{V}_{4}}{2\eta _{4}} -2V_{1}\,, \hspace{1cm} \gamma ^{2}=\sqrt{\frac{\lambda -1}{3\lambda -1}}\frac{\eta _{5}}{\eta _{4}} \frac{\tilde{V}_{4}}{\chi ^{2}}\,, \\
\delta ^{2}=\frac{\lambda -1}{3\lambda -1} \frac{1}{\chi ^{4}}\left[ (8g_{7}-3g_{8})+2\frac{\eta _{5}^{2}}{\eta _{4}} \right]\,,\hspace{1cm} \epsilon ^{2}=\frac{\lambda -1}{3\lambda -1} \frac{(8g_{2}-3g_{3})}{\chi ^{2}}\,,\hspace{1cm} \tilde{c}_{s}^{2}=\frac{\lambda -1}{3\lambda -1}\,,\hspace{1cm} \zeta ^{2}=\gamma ^{2}\,,
\label{coeff_uv}
\end{eqnarray}
\begin{equation}
\Xi _{2}=-\Omega _{2}=\sqrt{\frac{\lambda -1}{3\lambda -1}}\frac{\dot{\phi}_{0}}{2\chi } \left( \frac{(3\lambda -1)}{(\lambda -1)} +3+2\frac{\eta _{5}}{\eta _{4}} \right)\,,
\end{equation}
\begin{alignat}{1}
\Xi _{3}=  \sqrt{\frac{\lambda -1}{3\lambda -1}}  \frac{\ddot{\phi}_{0}}{\chi} \left( 3+2\frac{\eta _{5}}{\eta _{4}} \right) \,, \qquad \Omega _{3}= \sqrt{\frac{\lambda -1}{3\lambda -1}} \frac{\ddot{\phi}_{0}}{\chi}\,.
\end{alignat}

\appendix

\section*{Appendix 4 : Particular Solutions}
\label{sec:appendix_4}

\noindent{\bf Region I}:

We are going to evaluate the particular solution, $\delta u_{I}=\Gamma _{1}+\Gamma _{2}$. Making 
the change of variable $\bar{\eta}=k\eta$, we have
\begin{alignat}{1}
\Gamma_{1} & =\mp i\left(\frac{\pi}{2b}\right)^{3/2}\frac{1}{\sin\left(\pi/2b\right)}\frac{\bar{\eta}^{1/2}}{k^{1/2}}J_{1/2b}\left(z_{1}\right)\times \nonumber \\
 & \times\left\{ -\frac{\Xi_{2}\alpha}{bl_{0}}\left[e^{\pm ix_{i}}\int_{\bar{\eta}_{i}(k)}^{\bar{\eta}_{1}}d\bar{\eta}k^{\frac{3-b}{2}}\bar{\eta}^{\frac{1+b}{2}}J_{1/2b}\left(z\right)J_{-1/2b}\left(z\right)-e^{\pm iy_{i}}\int_{\bar{\eta}_{i}(k)}^{\bar{\eta}_{1}}d\bar{\eta}k^{\frac{3-b}{2}}\bar{\eta}^{\frac{1+b}{2}}J_{-1/2b}\left(z\right)J_{-1/2b}\left(z\right)\right]\right.  \nonumber \\
 & +\frac{\gamma^{2}}{l_{0}^{2}}\left[e^{\pm ix_{i}}\int_{\bar{\eta}_{i}(k)}^{\bar{\eta}_{1}}d\bar{\eta}k^{3-b}\bar{\eta}^{b}J_{1/2b}\left(z\right)J_{-1/2b}\left(z\right)-e^{\pm iy_{i}}\int_{\bar{\eta}_{i}(k)}^{\bar{\eta}_{1}}d\bar{\eta}k^{3-b}\bar{\eta}^{b}J_{-1/2b}\left(z\right)J_{-1/2b}\left(z\right)\right]  \nonumber \\
 & \left.+\Xi_{3}l_{0}^{2}\left[e^{\pm ix_{i}}\int_{\bar{\eta}_{i}(k)}^{\bar{\eta}_{1}}d\bar{\eta}k^{b-3}\bar{\eta}^{2-b}J_{1/2b}\left(z\right)J_{-1/2b}\left(z\right)-e^{\pm iy_{i}}\int_{\bar{\eta}_{i}(k)}^{\bar{\eta}_{1}}d\bar{\eta}k^{b-3}\bar{\eta}^{2-b}J_{-1/2b}\left(z\right)J_{-1/2b}\left(z\right)\right]\right\} 
\end{alignat}
 \begin{alignat}{1}
\Gamma_{2} & =\mp i\left(\frac{\pi}{2b}\right)^{3/2}\frac{1}{\sin\left(\pi/2b\right)}\frac{\bar{\eta}^{1/2}}{k^{1/2}}J_{-1/2b}\left(z_{1}\right)\times  \nonumber \\
 & \times\left\{ -\frac{\Xi_{2}\alpha}{bl_{0}}\left[e^{\pm ix_{i}}\int_{\bar{\eta}_{i}(k)}^{\bar{\eta}_{1}}d\bar{\eta}k^{\frac{3-b}{2}}\bar{\eta}^{\frac{1+b}{2}}J_{1/2b}\left(z\right)J_{1/2b}\left(z\right)-e^{\pm iy_{i}}\int_{\bar{\eta}_{i}(k)}^{\bar{\eta}_{1}}d\bar{\eta}k^{\frac{3-b}{2}}\bar{\eta}^{\frac{1+b}{2}}J_{1/2b}\left(z\right)J_{-1/2b}\left(z\right)\right]\right.  \nonumber \\
 & +\frac{\gamma^{2}}{l_{0}^{2}}\left[e^{\pm ix_{i}}\int_{\bar{\eta}_{i}(k)}^{\bar{\eta}_{1}}d\bar{\eta}k^{3-b}\bar{\eta}^{b}J_{1/2b}\left(z\right)J_{1/2b}\left(z\right)-e^{\pm iy_{i}}\int_{\bar{\eta}_{i}(k)}^{\bar{\eta}_{1}}d\bar{\eta}k^{3-b}\bar{\eta}^{b}J_{1/2b}\left(z\right)J_{-1/2b}\left(z\right)\right]  \nonumber \\
 & \left.+\Xi_{3}l_{0}^{2}\left[e^{\pm ix_{i}}\int_{\bar{\eta}_{i}(k)}^{\bar{\eta}_{1}}d\bar{\eta}k^{b-3}\bar{\eta}^{2-b}J_{1/2b}\left(z\right)J_{1/2b}\left(z\right)-e^{\pm iy_{i}}\int_{\bar{\eta}_{i}(k)}^{\bar{\eta}_{1}}d\bar{\eta}k^{b-3}\bar{\eta}^{2-b}J_{1/2b}\left(z\right)J_{-1/2b}\left(z\right)\right]\right\} 
\end{alignat}
where $z=\alpha k^{3-b}\bar{\eta}^{b}/bl_{0}^{2}$.
In the case of de Sitter inflation, this particular solution reduces to:
\begin{alignat}{1}
\Gamma_{1} & =\mp\frac{i}{2}\left(\frac{\pi}{6}\right)^{3/2}\frac{\bar{\eta}^{1/2}}{k^{1/2}}J_{1/2b}\left(z_{1}\right)\times  \nonumber \\
 & \times\left\{ -\frac{\Xi_{2}\alpha}{3l_{0}}\left[e^{\pm ix_{i}}\int_{\bar{\eta}_{i}(k)}^{\bar{\eta}_{1}}d\bar{\eta}\bar{\eta}^{2}J_{1/2b}\left(z\right)J_{-1/2b}\left(z\right)-e^{\pm iy_{i}}\int_{\bar{\eta}_{i}(k)}^{\bar{\eta}_{1}}d\bar{\eta}\bar{\eta}^{2}J_{-1/2b}\left(z\right)J_{-1/2b}\left(z\right)\right]\right. \nonumber \\
 & +\frac{\gamma^{2}}{l_{0}^{2}}\left[e^{\pm ix_{i}}\int_{\bar{\eta}_{i}(k)}^{\bar{\eta}_{1}}d\bar{\eta}\bar{\eta}^{3}J_{1/2b}\left(z\right)J_{-1/2b}\left(z\right)-e^{\pm iy_{i}}\int_{\bar{\eta}_{i}(k)}^{\bar{\eta}_{1}}d\bar{\eta}\bar{\eta}^{3}J_{-1/2b}\left(z\right)J_{-1/2b}\left(z\right)\right]  \nonumber \\
 & \left.+\Xi_{3}l_{0}^{2}\left[e^{\pm ix_{i}}\int_{\bar{\eta}_{i}(k)}^{\bar{\eta}_{1}}d\bar{\eta}\frac{1}{\bar{\eta}}J_{1/2b}\left(z\right)J_{-1/2b}\left(z\right)-e^{\pm iy_{i}}\int_{\bar{\eta}_{i}(k)}^{\bar{\eta}_{1}}d\bar{\eta}\frac{1}{\bar{\eta}}J_{-1/2b}\left(z\right)J_{-1/2b}\left(z\right)\right]\right\} 
 \label{gamma_1_b3}
\end{alignat}
\begin{alignat}{1}
\Gamma_{2} & =\mp\frac{i}{2}\left(\frac{\pi}{6}\right)^{3/2}\frac{\bar{\eta}^{1/2}}{k^{1/2}}J_{-1/2b}\left(z_{1}\right)\times  \nonumber \\
 & \times\left\{ -\frac{\Xi_{2}\alpha}{3l_{0}}\left[e^{\pm ix_{i}}\int_{\bar{\eta}_{i}(k)}^{\bar{\eta}_{1}}d\bar{\eta}\bar{\eta}^{2}J_{1/2b}\left(z\right)J_{1/2b}\left(z\right)-e^{\pm iy_{i}}\int_{\bar{\eta}_{i}(k)}^{\bar{\eta}_{1}}d\bar{\eta}\bar{\eta}^{2}J_{1/2b}\left(z\right)J_{-1/2b}\left(z\right)\right]\right.  \nonumber \\
 & +\frac{\gamma^{2}}{l_{0}^{2}}\left[e^{\pm ix_{i}}\int_{\bar{\eta}_{i}(k)}^{\bar{\eta}_{1}}d\bar{\eta}\bar{\eta}^{3}J_{1/2b}\left(z\right)J_{1/2b}\left(z\right)-e^{\pm iy_{i}}\int_{\bar{\eta}_{i}(k)}^{\bar{\eta}_{1}}d\bar{\eta}\bar{\eta}^{3}J_{1/2b}\left(z\right)J_{-1/2b}\left(z\right)\right]  \nonumber \\
 & \left.+\Xi_{3}l_{0}^{2}\left[e^{\pm ix_{i}}\int_{\bar{\eta}_{i}(k)}^{\bar{\eta}_{1}}d\bar{\eta}\frac{1}{\bar{\eta}}J_{1/2b}\left(z\right)J_{1/2b}\left(z\right)-e^{\pm iy_{i}}\int_{\bar{\eta}_{i}(k)}^{\bar{\eta}_{1}}d\bar{\eta}\frac{1}{\bar{\eta}}J_{1/2b}\left(z\right)J_{-1/2b}\left(z\right)\right]\right\} 
 \label{gamma_2_b3}
\end{alignat}
where $z=\alpha \bar{\eta}^{3}/3l_{0}^{2}$.

We can now evaluate the 12 integrals that we have in this solution. We are only computing them 
for de Sitter inflation. The solution of these integrals are polynomials multiplied by generalized 
hypergeometric functions. This functions can be written in terms of Meijer functions, a general 
function that contains most of the special functions as a special case and that it is defined as a 
path integral in the complex plane \cite{hyper,gauge_trans}. For example, the integral:
\begin{align}
i_{1}=i_{8}&=\int_{\bar{\eta}_{i}(k)}^{\bar{\eta}_{1}}d\bar{\eta}\bar{\eta}^{2}J_{-1/2b}\left(z\right)J_{1/2b}\left(z\right) \\
=&\frac{\bar{\eta}^{3}}{3}\Hypergeometric{2}{3}{1/2,1/2}{5/6,7/6,9/6}{-z^{2}}
=\frac{\bar{\eta}^{3}}{6\sqrt{\pi}}\MeijerG{1/2,1/2}{}{0}{1/6,-1/6,1/2}{z^{2}} 
\end{align}
All integrals give the generalized hypergeometric function with  $p=2$ and $q=p+1=3$ which 
is a special case of the Meijer function with $m=1$, $n=2=p$ and $q=4$, except for the last 
integral in (\ref{gamma_2_b3}) (and its counterpart in (\ref{gamma_1_b3})) that gives a  
hypergeometric function with $p=3$ and $q=4$ that it is a special case of a Meijer function 
with $m=1$, $n=3=p$ and $q=5$. 

To be able to work with these results we make use of the asymptotic expansion of the Meijer 
function for large $z$. For the Meijer functions present here, we use Theorem 1.8.5, the 
case (1.8.13) in \cite{hyper}, verifying that the functions obtained obey all the conditions 
imposed in the theorem:
\begin{equation}
\MeijerG[a,b]{m}{p}{n}{q}{z^{2}}  \sim D_{p,q}^{m,n} (\lambda ) H_{p,q} \left[z^{2} \exp [i\pi (\mu ^{*} -2\lambda )]\right]\,,
\end{equation}
where, $\mu ^{*}=q-m-n$, $\nu ^{*}=-p+m+n$ and $arg(z^{2})=0$. In all of our cases 
$\mu ^{*}=1=\mu ^{*}$ and 
$ D_{p,q}^{m,n} (\lambda ) = A^{m,n}_{q} = (2\pi i)^{\mu *} \exp \left[ i\pi \left( \sum _{j=1}^{n} a_{j}- \sum _{j=m+1}^{q} b_{j} \right) \right]$ 
as follows from (1.7.23) and (1.7.2) of \cite{hyper}. The function $H{p,q}$ can be written as a series:
\begin{equation}
H_{p,q} [x]=\exp \left[ (p-q)x^{\frac{1}{p-q}}\right] x^{\rho ^{*}} \left[ \frac{(2\pi)^\frac{q-p-1}{2}}{(q-p)^1/2} +\mathcal{O}(x^{-\frac{1}{q-p}})\right]\,,
\end{equation}
with $\rho ^{*}=\left[  \sum _{j=m+1}^{q} b_{j} - \sum _{j=1}^{n} a_{j} +( p-q+1)/2\right]/(q-p)$.

This asymptotic expansion was applied in all of the results and, after some algebra, the particular 
solution simplifies to:
\begin{alignat}{1}
\Gamma_{1}+\Gamma_{2} & =\mp\frac{1}{4}\frac{1}{\left(c_{s}k\right)^{1/2}}\left(\frac{l_{0}^{2}}{\alpha}\right)^{2}\left\{ -\frac{\Xi_{2}\alpha}{2l_{0}}\left(\frac{1}{\bar{\eta}_{1}^{3}}e^{-2iz_{1}}-\frac{1}{\bar{\eta}_{i}^{3}}e^{-2iz_{i}}\right)\left(\cos\left(z_{1}-\frac{\pi}{3}\right)c_{1}^{-}+\cos\left(z_{1}-\frac{\pi}{6}\right)c_{2}^{-}\right)+\right. \nonumber \\
 & \left.+\left[\frac{\gamma^{2}}{l_{0}^{2}}\left(\frac{1}{\bar{\eta}_{1}^{2}}e^{-2iz_{1}}-\frac{1}{\bar{\eta}_{i}^{2}}e^{-2iz_{i}}\right)-\Xi_{3}l_{0}^{2}\left(\frac{1}{\bar{\eta}_{1}^{6}}e^{-2iz_{1}}-\frac{1}{\bar{\eta}_{i}^{6}}e^{-2iz_{i}}\right)\right]\left(\cos\left(z_{1}-\frac{\pi}{3}\right)c_{1}^{+}+\cos\left(z_{1}-\frac{\pi}{6}\right)c_{2}^{+}\right)+\right. \nonumber \\
 & \left. +4\Xi_{3}l_{0}^{2}\left(\frac{l_{0}^{2}}{\alpha}\right)^{-2}\ln\left[\left(\frac{\lambda\left(\eta_{i}\right)}{l_{c}}\right)^{3}\right]\left(\cos\left(z_{1}-\frac{\pi}{3}\right)+\cos\left(z_{1}-\frac{\pi}{6}\right)\right)\right\} \,.
 \label{sol_part_I}
\end{alignat}

\noindent{\bf Region II}:

To determine the particular solution, $\delta u_{II}=\tilde{\Gamma} _{1}+\tilde{\Gamma} _{2}$, we 
make the change of variable $\bar{\eta}=k\eta$, and get:
\begin{alignat}{1}
\delta u_{II} & =-\frac{i}{2}\left\{ -i\:\Xi_{2}l_{0}\left[\left(\tilde{B}_{1}e^{ic_{s}\bar{\eta}_{2}}-\tilde{B}_{2}e^{-ic_{s}\bar{\eta}_{2}}\right)\int_{\bar{\eta}_{1}}^{\bar{\eta}_{2}}d\bar{\eta}k^{\frac{b-3}{2}}\bar{\eta}^{\frac{1-b}{2}}+\int_{\bar{\eta}_{1}}^{\bar{\eta}_{2}}d\bar{\eta}k^{\frac{b-3}{2}}\bar{\eta}^{\frac{1-b}{2}}\left(\tilde{B}_{1}e^{-ic_{s}\bar{\eta}_{2}}e^{2ic_{s}\bar{\eta}}-\tilde{B}_{2}e^{ic_{s}\bar{\eta}_{2}}e^{-2ic_{s}\bar{\eta}}\right)\right]\right. \nonumber \\
 & +\frac{\gamma^{2}}{l_{0}^{2}}\left[\left(\tilde{B}_{1}e^{ic_{s}\bar{\eta}_{2}}+\tilde{B}_{2}e^{-ic_{s}\bar{\eta}_{2}}\right)\int_{\bar{\eta}_{1}}^{\bar{\eta}_{2}}d\bar{\eta}\, k^{3-b}\bar{\eta}^{b-1}+\int_{\bar{\eta}_{1}}^{\bar{\eta}_{2}}d\bar{\eta}\, k^{3-b}\bar{\eta}^{b-1}\left(\tilde{B}_{1}e^{-ic_{s}\bar{\eta}_{2}}e^{2ic_{s}\bar{\eta}}+\tilde{B}_{2}e^{ic_{s}\bar{\eta}_{2}}e^{-2ic_{s}\bar{\eta}}\right)\right] \nonumber \\
 & \left.+\Xi_{3}l_{0}^{2}\left[\left(\tilde{B}_{1}e^{ic_{s}\bar{\eta}_{2}}+\tilde{B}_{2}e^{-ic_{s}\bar{\eta}_{2}}\right)\int_{\bar{\eta}_{1}}^{\bar{\eta}_{2}}d\bar{\eta}k^{b-3}\bar{\eta}^{1-b}+\int_{\bar{\eta}_{1}}^{\bar{\eta}_{2}}d\bar{\eta}k^{b-3}\bar{\eta}^{1-b}\left(\tilde{B}_{1}e^{-ic_{s}\bar{\eta}_{2}}e^{2ic_{s}\bar{\eta}}+\tilde{B}_{2}e^{ic_{s}\bar{\eta}_{2}}e^{2ic_{s}\bar{\eta}}\right)\right]\right\} \,,
\end{alignat}
where $\tilde{B}_{1}$ and $\tilde{B}_{2}$ are the coefficients of the homogeneous equation in 
Region II for $v_{k}$. They are equal to (\ref{B}), since the coefficients of the equations of motions 
of $u$ and $v$ were set to be the same, for simplicity.

For de Sitter inflation, the solution is:
\begin{alignat}{1}
\delta u_{II} & =-\frac{i}{2}\left\{ -i\:\Xi_{2}l_{0}\left[\left(\tilde{B}_{1}e^{ic_{s}\bar{\eta}_{2}}-\tilde{B}_{2}e^{-ic_{s}\bar{\eta}_{2}}\right)\int_{\bar{\eta}_{1}}^{\bar{\eta}_{2}}d\bar{\eta}\frac{1}{\bar{\eta}}+\int_{\bar{\eta}_{1}}^{\bar{\eta}_{2}}d\bar{\eta}\frac{1}{\bar{\eta}}\left(\tilde{B}_{1}e^{-ic_{s}\bar{\eta}_{2}}e^{2ic_{s}\bar{\eta}}-\tilde{B}_{2}e^{ic_{s}\bar{\eta}_{2}}e^{-2ic_{s}\bar{\eta}}\right)\right]\right. \nonumber \\
 & +\frac{\gamma^{2}}{l_{0}^{2}}\left[\left(\tilde{B}_{1}e^{ic_{s}\bar{\eta}_{2}}+\tilde{B}_{2}e^{-ic_{s}\bar{\eta}_{2}}\right)\int_{\bar{\eta}_{1}}^{\bar{\eta}_{2}}d\bar{\eta}\,\bar{\eta}^{2}+\int_{\bar{\eta}_{1}}^{\bar{\eta}_{2}}d\bar{\eta}\,\bar{\eta}^{2}\left(\tilde{B}_{1}e^{-ic_{s}\bar{\eta}_{2}}e^{2ic_{s}\bar{\eta}}+\tilde{B}_{2}e^{ic_{s}\bar{\eta}_{2}}e^{-2ic_{s}\bar{\eta}}\right)\right] \nonumber \\
 & \left.+\Xi_{3}l_{0}^{2}\left[\left(\tilde{B}_{1}e^{ic_{s}\bar{\eta}_{2}}+\tilde{B}_{2}e^{-ic_{s}\bar{\eta}_{2}}\right)\int_{\bar{\eta}_{1}}^{\bar{\eta}_{2}}d\bar{\eta}\frac{1}{\bar{\eta}^{2}}+\int_{\bar{\eta}_{1}}^{\bar{\eta}_{2}}d\bar{\eta}\frac{1}{\bar{\eta}^{2}}\left(\tilde{B}_{1}e^{-ic_{s}\bar{\eta}_{2}}e^{2ic_{s}\bar{\eta}}+\tilde{B}_{2}e^{ic_{s}\bar{\eta}_{2}}e^{-2ic_{s}\bar{\eta}}\right)\right]\right\} \,, .
 \label{delta_u_II_b3}
\end{alignat}


\end{document}